\documentclass[lettersize,journal]{IEEEtran}

\usepackage{amsmath,graphicx}
\usepackage{amsfonts}
\usepackage{amssymb}
\usepackage{epsfig}
\usepackage{mathrsfs}
\usepackage{graphicx}
\usepackage{algorithm}
\usepackage{epstopdf}
\usepackage{balance}
\usepackage{color}
\usepackage{cite}
\usepackage{algorithmic}
\usepackage{array}
\usepackage{textcomp}
\usepackage{stfloats}
\usepackage{verbatim}
\usepackage{bbding}
\usepackage{pifont}
\usepackage{bbm}
\usepackage{siunitx}
\usepackage{caption}
\usepackage{subcaption}

\def\lf{\left\lfloor}   
\def\rf{\right\rfloor}

\usepackage[nocomma]{optidef}

\DeclarePairedDelimiter\floor{\lfloor}{\rfloor}

\newtheorem{lemma}{Lemma}[section]

\begin{document}
% To control the max number of authors use the following line
\bstctlcite{IEEEexample:BSTcontrol}
% The bib file should contai something like this
%@IEEEtranBSTCTL{IEEEexample:BSTcontrol,
%CTLuse_forced_etal       = "yes",
%CTLmax_names_forced_etal = "3",
%CTLnames_show_etal       = "2" }

\title{Multi-user Goal-oriented Communications with Energy-efficient Edge Resource Management}

\author{Francesco Binucci$^1$, Paolo Banelli$^1$,~\IEEEmembership{Member,~IEEE}, \\Paolo~Di Lorenzo$^2$,~\IEEEmembership{Senior member,~IEEE}, and Sergio Barbarossa$^2$,~\IEEEmembership{Fellow,~IEEE}
\\
\smallskip
$^1$ University of Perugia, Via G. Duranti 93, 06125, Perugia, Italy;\\
$^2$ Sapienza University of Rome, Via Eudossiana 18, 00184, Rome, Italy;\\ 
\smallskip
Email: \texttt{ \{francesco.binucci, paolo.banelli\}@unipg.it, \{paolo.dilorenzo, sergio.barbarossa\}@uniroma1.it} \vspace{-.5cm}
%\thanks{Binucci and Banelli are with the Dept. of Engineering, University of Perugia, Via G. Duranti 93, 06125, Perugia, Italy; Email: \texttt{francesco.binucci@studenti.unipg.it, paolo.banelli@unipg.it}.\newline \indent Di Lorenzo and Barbarossa are with the Dept. of Information Engineering, Electronics, and Telecommunications, Sapienza University of Rome, Via Eudossiana 18, 00184, Rome, Italy; E-mail: \texttt{sergio.barbarossa@uniroma1.it}. This work has been supported by by Italian Minister o University and Research under the PRIN 2017 Liquid Edge Contract.}
\thanks{This work has been supported by MIUR under the PRIN 2017 Liquid Edge Contract. The work of P. Di Lorenzo and S. Barbarossa was funded by the SNS-JU-2022 project ADROIT6G under agreement n. 101095363. This work was partially supported by the European Union under the Italian National Recovery and Resilience Plan (NRRP) of NextGenerationEU, partnership on “Telecommunications of the Future” (PE00000001 - program “RESTART”).}
}

%\thanks{footnotes }}
%% <-this % stops a space
%\thanks{Manuscript received April 19, 2021; revised August 16, 2021.}}
%
%% <-this % stops a space
%\thanks{This paper was produced by the IEEE Publication Technology Group. They are in Piscataway, NJ.}

%\markboth{Journal of \LaTeX\ Class Files,~Vol.~XX, No.~XX, Month~2022}%
%{Shell \MakeLowercase{\textit{et al.}}: Edge-assisted Goal-oriented Wireless Communications}

%\IEEEpubid{0000--0000/00\$00.00~\copyright~2022 IEEE}

\maketitle

\vspace{-.3cm}

\begin{abstract}
Edge Learning (EL) pushes the computational resources toward the edge of 5G/6G network to assist mobile users requesting delay-sensitive and energy-aware  intelligent services. A common challenge in running inference tasks from remote is to extract and transmit only the features that are most significant for the inference task. From this perspective, EL can be effectively coupled with goal-oriented communications, whose aim is to transmit only the information {\it relevant} to perform the inference task, under prescribed accuracy, delay, and energy constraints.
In this work, we consider a multi-user/single server wireless network, where the users can opportunistically decide whether to perform the inference task by themselves or, alternatively, to offload the data to the edge server for remote processing. The data to be transmitted undergoes a goal-oriented compression stage performed using a convolutional encoder, jointly  trained with a convolutional decoder running at the edge-server side.
Employing Lyapunov optimization, we propose a method to jointly and dynamically optimize the selection of the most suitable encoding/decoding scheme, together with the allocation of computational and transmission resources, across all the users and the edge server.
%More specifically, we propose two strategies: i) a minimum-energy allocation, under delay and accuracy constraints,  and ii) a maximum-accuracy strategy under latency and energy guarantees. 
Extensive simulations confirm the effectiveness of the proposed approaches and highlight the trade-offs between energy, latency, and learning accuracy. 
\end{abstract}


\section{Introduction}
\IEEEPARstart{T}{he} advent of the fifth/sixth generation of mobile communications has radically changed the network concept, from a pure communication infrastructure to a key enabler for pervasive services, which are highly based on Artificial Intelligence (AI) and Machine Learning (ML).Typical examples can be found in augmented reality, autonomous driving, massive Internet of Things, and mission critical applications \cite{jiang2021road}.
In these scenarios, the service delay and the  reliability constraints are often very restrictive, and this motivates the need to design a holistic system where communication, computation, learning, and control are jointly managed in order to reach reliability, energy efficiency, and sustainability.

The need to process a huge amount of data, in real-time, through proper AI/ML techniques, has driven researchers to design training/inference tasks at the wireless edge, in collective as well as distributed fashions. This has led to the definition of the so called Edge Intelligence (EI) paradigm \cite{zhou2019edge}. In this view, the allocation of system resources
% according to specific optimization strategies, 
in order to reach prescribed target performance in terms of latency, accuracy, and energy consumption has been already considered in \cite{park2019wireless, wang2018edge, zhu2020toward,merluzzi2021wireless}. Specifically, EI allows User Equipments (UEs) connected to a mobile network to opportunistically offload their learning tasks to Edge Servers (ESs), which are placed in the edge, nearby the Radio Access Points (RAPs). This allows the efficient management of system resources, such as transmission rate, bandwidth, and CPU clock rates, according to specific optimization strategies, which are mainly focused on the trade-offs between energy consumption, overall latency, and learning accuracy \cite{merluzzi2021wireless}.
%\textcolor{red}{The resource optimization can be tackled by a plethora of mathematical tools, such as Stochastic Optimization, Game Theory, Deep Reinforcement Learning \cite{li2018deep}, etc..}

Clearly, in a resource optimization perspective, it would be useful to offload to the ESs only the (minimum) amount of information strictly necessary to fulfill the learning task with the desired accuracy, while respecting the performance requirements. This intuitive consideration, jointly with the huge increase of traffic envisaged in future 6G networks \cite{giordani2020toward}, motivates the search for a new communication paradigm, alternative to the classical Shannon design. In this view, a valuable candidate is represented by Goal-Oriented Communications (GOC) \cite{strinati20216g}. More specifically, if the goal of communication is to perform an inference task on the data collected by the UE,  rather than requiring the accurate reproduction of all the transmitted bits at the receiver side, the aim of GOC is to transmit only the information that is most relevant to run the inference task at the ES, guaranteeing a prescribed level of decision accuracy and system performance. In this way, it is possible to help the UEs to save transmission resources and avoid unnecessary data rate growth, still respecting application constraints, such as service delay and energy consumption. 

\textbf{Related works.} 
Seminal EI frameworks, with a wireless offloading strategy, have been proposed in \cite{merluzzi2020dynamic, merluzzi2021wireless}, which save transmission resources by simply allocating, in a dynamic fashion, the number of (quantization) bits used by UEs to transmit their data to the ES. This compression strategy has also been employed in \cite{merluzzi2022energy} and \cite{merluzzi2021ensemble}, where edge classification and ensemble learning are considered, respectively, with reliability guarantees.
% An alternative to the conventional separate design of source and channel encoders has been proposed in  \cite{bourtsoulatze2019deep}, where deep neural networks (DNN) have been exploited to realize joint source/channel coding (JSCC), which grants superior image retrivial performance in a finite block-length regime.
% JSCC was investigated also in \cite{lee2019deep}, focusing directly on the recognition accuracy rather than on image reconstruction followed by a separate classification network. An image retrieval scheme was proposed in \cite{jankowski2020wireless} that, rather than sending the image, directly maps the image feature vectors on channel input symbols, while the server retrieves the images from the noisy channel output, without exploiting any classical channel code.
A more principled data reduction strategy, better matched to the learning task and based on the Information Bottleneck (IB) \cite{tishby2000information} \cite{goldfeld2020information}, has been proposed in \cite{ICASSPBottleneck}. However, the IB principle admits a closed form solution for the encoder only if the overall statistics are jointly Gaussian \cite{chechik2005information} \cite{ICASSPBottleneck}, or a solution achievable through an iterative mechanism, if the statistics are discrete. When the sensed data and decision outputs are neither jointly Gaussian, nor discrete with manageable cardinality, it is not easy to derive the IB solution and the source encoding problem can be reformulated using the so called variational IB (VIB), as recently explored in \cite{shao2021learning} and in \cite{shao2022task}, where a cooperative (multi-device) inference framework is proposed.

A possibility to further deviate from the classical communication design is offered by Joint Source Channel/Coding (JSCC), which has received increasing attention with the wide spread use of Deep Neural Networks (DNNs). Quite recently, several works have proposed to replace the classical cascade of source and channel encoders with a DNN properly trained with respect to the specific task. For instance, \cite{bourtsoulatze2019deep} proposed a DNN-based JSCC scheme to achieve higher performance in finite block-length regime for image retrieval applications. Furthermore, if the task of communication is image recognition, it makes sense to design the JSCC architecture directly focusing on the learning task, rather than on the image reconstruction followed by the recognition task, as proposed in \cite{lee2019deep}. 
The authors of \cite{jankowski2020wireless} presented a scheme for image retrieval where the extracted vector features are directly mapped to the channel input symbols, without resorting to any channel coding technique, and  the server retrieves the most relevant images directly from the noisy channel output. This approach has been extended in \cite{tung2022deepjscc}, where the extracted features are quantized before being mapped onto the channel symbols.
In \cite{yang2022ofdm}, JSCC is coupled with an OFDM system operating over a frequency-selective channel, while \cite{dai2022nonlinear} considers the combination of JSCC with non-linear transform coding (NTC) \cite{balle2020nonlinear}.

As far as goal-oriented (also known as task-oriented) communications is concerned, several recent works testify the emerging relevance of this topic. For instance, in \cite{farshbafan2021common} and \cite{farshbafan2022curriculum} GOCs have been exploited to define the \emph{common-language} between a listener and a speaker, employing Reinforcement Learning (RL) and Curriculum Learning (CL), while a transformer-based approach has been proposed to assist image and text transmissions \cite{xie_letaief2021task}.
A noise-aware JSCC for text-transmission is described and assessed in \cite{peng2022robust}, while \cite{jiang2022deep} exploited a hybrid automatic repeat request (HARQ) scheme to improve reliability in sentence semantic transmission.
Other examples of image classification for Unmanned Aerial Vehicle (UAV) applications, and a GOC-assisted Visual Question Answering (VQA) task, can be found in \cite{kang2021task} and \cite{xie2021task}, respectively.
Furthermore, \cite{lee2019deep} and \cite{jankowski2020wireless} motivate the use of GOC schemes for computer vision applications, by showing the accuracy improvements they provide in image-classification and re-identification tasks of humans and cars, respectively. Finally, the impact of goal-oriented communications has also been analyzed in speech recognition tasks \cite{weng2021semantic}.

However, none of the works cited above considered the dynamic optimization of the data reduction strategy for multi-user goal-oriented communications, {\it jointly} with the global network resource management, under {\it prescribed performance guarantees}, as we do in this manuscript. Along this line, in \cite{jasp_sumbitted_binucci} we proposed  minimum-energy and maximum-accuracy resource allocation strategies for edge-assisted image classification tasks, in a \emph{single user}/single server scenario, whereas in \cite{eusipco_paper} we reported some preliminary results on the extension to the multi-user scenario, which we will further develop and investigate more thoroughly hereinafter.

\textbf{Our contributions.} The main contributions of this work concern the system architecture, the optimization strategies, and the simulation results. They can be summarized as follows:
\subsubsection{System Architecture} Extending the preliminary strategies presented in \cite{eusipco_paper}, we consider a \emph{multi-user} goal-oriented communication scenario, where {\it multiple} UEs 
%({\it differently from \cite{eusipco_paper}}) 
may decide to offload their learning tasks to an ES  (\textit{or not}). 
Each user relies on a bank of source encoders, each one associated to a specific compression ratio, which dynamically compresses the data-units (DUs) to be transmitted to the ES, depending on the online system state. Specifically, exploiting convolutional encoders (CEs), i.e., the encoders of convolutional auto-encoders (CAE), as in  \cite{jasp_sumbitted_binucci}, we improve their performance by a new training function. %However, within the proposed framework, we may exploit any possible alternative technique to smartly reduce the data to be transmitted.
The ES carries out multiple, user-independent, inference tasks, using a bank of convolutional classifiers (CCs), i.e., CNNs, each one matched to the CE used at the UE. The overall CE-CC structure is instrumental to {\it split the classification task between UE and ES}.
\subsubsection{Optimization Strategies:} We implement a {\it dynamical split} of the inference task, selecting, in each time slot,  the most suitable pair of CE-CCs, within the bank of available (pre-trained) CE-CCs, depending on the channel state and on the online accuracy and performance. More specifically, resorting to Lyapunov optimization, we implement a multi-user \textit{dynamical} goal-oriented source compression architecture that selects the CE-CC pair and allocates computational and communication resources, trading off energy consumption (including both UEs and ES), delay and classification accuracy. Hereinafter, we extend the preliminary results and optimization strategy shown in \cite{eusipco_paper}, by considering also a multi-user Maximum Accuracy strategy, with guaranteed (maximum) Delay bounds and Energy consumption (MADE). Furthermore, we let every UE able to decide whether to perform the inference task locally or to offload it to the ES, since there might be applications where the UE hardware is capable of running the application locally, or it could be more convenient, for the overall resource management, to do that.
\subsubsection{Simulation Scenarios} We investigate scenarios that were not analyzed in \cite{eusipco_paper}, where each UE has different service requirements and constraints. The wide set of possible scenarios, optimization strategies, and simulation results, significantly extends the results in \cite{eusipco_paper}, highlighting the effectiveness and flexibility of the proposed holistic resource management.

\textbf{Outline}. The paper is organized as follows. Sec. \ref{sec:training} illustrates the goal-oriented communication system and the related joint training procedure of both the CEs and the CCs for classification purposes. Sec. \ref{sec:system_model} describes the overall system model used in the formulation of the resource optimization strategies, which are then solved in Sec. \ref{sec:opt_strat_develop} exploiting stochastic Lyapunov optimization. In Sec. \ref{sec:sim_res} we discuss our experimental results and, finally, in Sec. \ref{sec:conclusion_future} we draw some conclusions and highlight future research directions.

%\section{Training procedure}\label{sec:training}
%The training procedure for the classification network we are going to describe, is tailored to the goal-oriented communication system foreseen in this manuscript:

\section{Classification network and training}\label{sec:training}
\begin{figure}[t]
\centering
\includegraphics[width=1.00\linewidth]{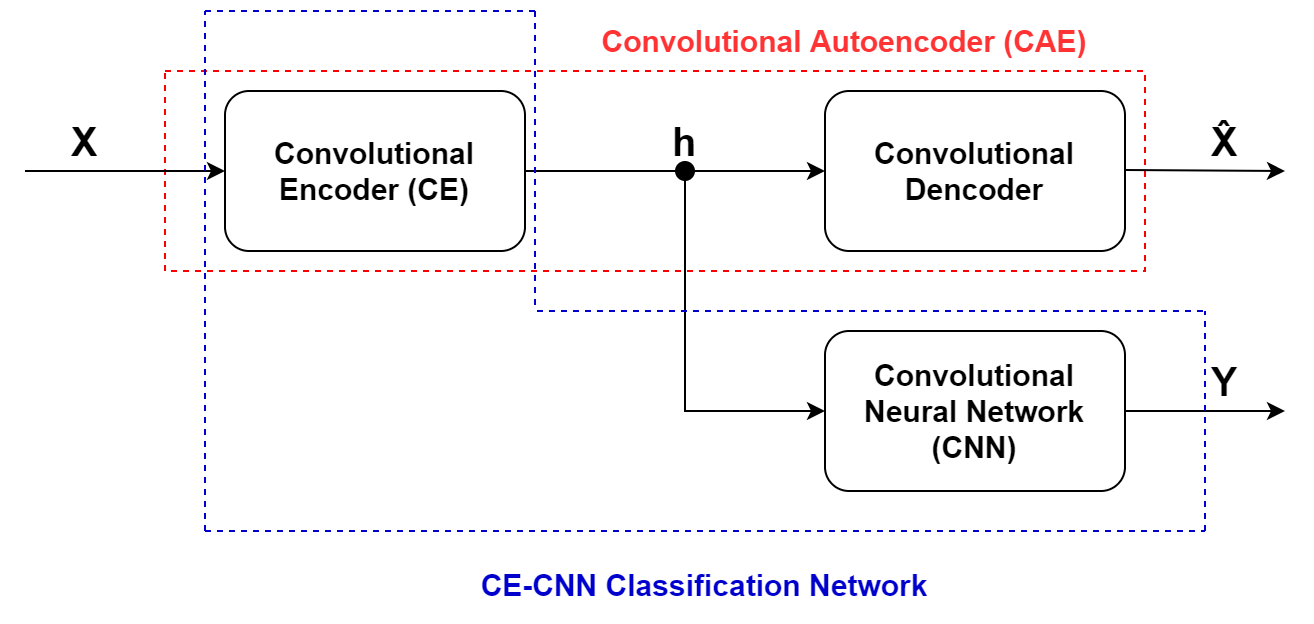}
    \caption{Training scheme: the output of the CE $h$ feeds both the ES classification CNN and a CD.}
    \label{fig:chained-scheme}
\end{figure}

This section describes the architecture employed to make parsimonious use of transmission energy and bandwidth. Specifically, we compress the UEs data-units (DUs) (i.e., the input of the learning task), before they are transmitted to the ES. The latter has to perform the learning task without sacrificing a prescribed target accuracy.
As more deeply explained in \cite{jasp_sumbitted_binucci}, the Information Bottleneck (IB) \cite{tishby2000information} is a promising theoretical framework to meaningfully compress the data-source in a goal-oriented perspective. However, IB admits a closed form solution only when the associated statistics are discrete or Gaussian distributed \cite{gaussian_ib} \cite{ICASSPBottleneck}.
Thus, since in the multi-class image classification task we are focusing on, the  Gaussian assumptions do not hold true and a meaningful definition of mutual information is problematic \cite{larkin2016reflections}, we proposed in \cite{jasp_sumbitted_binucci} a heuristic approximation of the IB that nicely fits with our goal-oriented strategy. Specifally, our approach is based on the deployment of a tunable data-compression at the UEs that is useful for the associated inference task at the ES. Without restriction of generality for the overall GOCs architecture and its resource management, we resort to banks of CEs to compress images at the UE side, according to a layer-by-layer max-pooling strategy. The CEs are coupled with CCs at the ES to perform the final decision, as summarized in Fig. \ref{fig:chained-scheme} for a single UE.

 % Side by side figures 

As detailed in \cite{jasp_sumbitted_binucci}, a CE may be realized as:
\begin{itemize}
\item \textbf{Short-CE:} 
It resizes the images to the desired resolution by a single convolutional layer followed by a max-pooling layer.
\item \textbf{Deep-CE:} It down-samples the images by multiple convolutional layers, each one followed by a max-pooling layer that  halves the size of the (pseudo) image.
\end{itemize}
Note that our goal is to classify the images and not to reproduce them. Thus, for the CE-CCs compression and classification network shown in  Fig.\ref{fig:chained-scheme}, we have to consider a different learning cost function  than those used for classical CAEs.
Specifically, we resort to the following objective function
%\small
\begin{mini}[2]
{\theta,\phi}{\frac{1}{N_t}\sum_{n=1}^{N_t} L_{ce}(Y_{n},\hat{Y}_{n},\phi,\theta)+\lambda L_{mse}(X_{n},\hat{X}_{n},\theta),}{}{}
\label{optim:regularized-training}
\end{mini}
%\normalsize
where $L_{ce}(Y_{n},\hat{Y}_{n},\phi,\theta)$ is the cross-entropy loss, used in order to control the performance of the ES classification task, while $L_{mse}(X_{n},\hat{X}_{n},\theta)$ is the \emph{Mean Squared Error} between the input and the reconstructed version $\hat{X}$ of the full CAE.
%Note that, differently from what we proposed in  \cite{jasp_sumbitted_binucci},
Note that the cross-entropy loss in \eqref{optim:regularized-training} is a proxy of the mutual information $I(h;Y)$ \cite{boudiaf2020unifying}.
Thus, minimizing the cross-entropy, we maximize the $I(h;Y)$ for a fixed CE architecture (compression size) and this constitutes the link of the proposed approach with the IB principle.
However, differently from what we did in \cite{jasp_sumbitted_binucci}, \eqref{optim:regularized-training} considers also the output MSE of a Convolutional Decoder (CD), i.e., that part of the CAE that is typically used for image reconstruction.
Actually, the presence in \eqref{optim:regularized-training} of this (regularizing) MSE penalty term favours a meaningful feature extraction \cite{NEURIPS2018_2a38a4a9}, which can improve the performance of the overall learning task, for proper values of the parameter $\lambda$.
Anyway, note that the CD is taken into account only during the CE-CCs training, while it is not used for classification, as clarified by Fig. \ref{fig:chained-scheme}.
Each CE-CC pair has to be properly trained, possibly off-line, by a third party. Thus, although it would be interesting to analyze how to train the classification network by the same wireless edge-computing architecture we consider herein for classification, this is not the object of this manuscript and is left for future studies.

% \textbf{Remark.} In our goal-oriented scenario, any specific architecture of the CE at the UE, fixes the compression degree (i.e., the size of the compressed pseudo-images $h$), which may be not optimal: this is the main point of divergence between the original IB formulation and our compression framework. However, there exists an important link with the IB: indeed, the cross-entropy loss in \eqref{optim:regularized-training} can be seen as a proxy of the mutual information $I(h;Y)$ \cite{boudiaf2020unifying}. Thus, minimizing the cross-entropy, we maximize the $I(h;Y)$ for a fixed CE architecture (compression size).

\textbf{JPEG compression.} Targeting good classification performance, the CE compresses the images by a down-sampling principle, due to the max-pooling strategy at each layer. However, this design does not take into account the wireless communication between UEs and ES. Thus, while the size of the latent representation $h$ of a CE output (see Fig. \ref{fig:chained-scheme}) may be optimal for a target classification accuracy, it could be still sub-optimal with respect to the file size of the compressed data-units, leading to huge costs in terms of transmission energy and time. This problem justifies the employment of a further zipping (compression) phase on $h$, before transmitting it to the ES, which will unzip it back to $h$ at the CC input. Due to the nature of the classification task and the structure of the pseudo-images $h$ extracted by the CE, we base this further compression at the UE on a JPEG codec, which proved to effectively reduce the file size of the data units, paying a reasonable price in terms of additional computational overhead from the UE perspective. The choice of JPEG is justified since it is a widely used zipping system, with a plethora of efficient implementations. Furthermore, despite its lossy nature, it has been proved that JPEG codecs do not significantly affect the classification performance of CNNs \cite{dodge2016understanding}.

\section{System model}\label{sec:system_model}

\begin{figure*}[t]
    \centering \includegraphics[width=\linewidth]{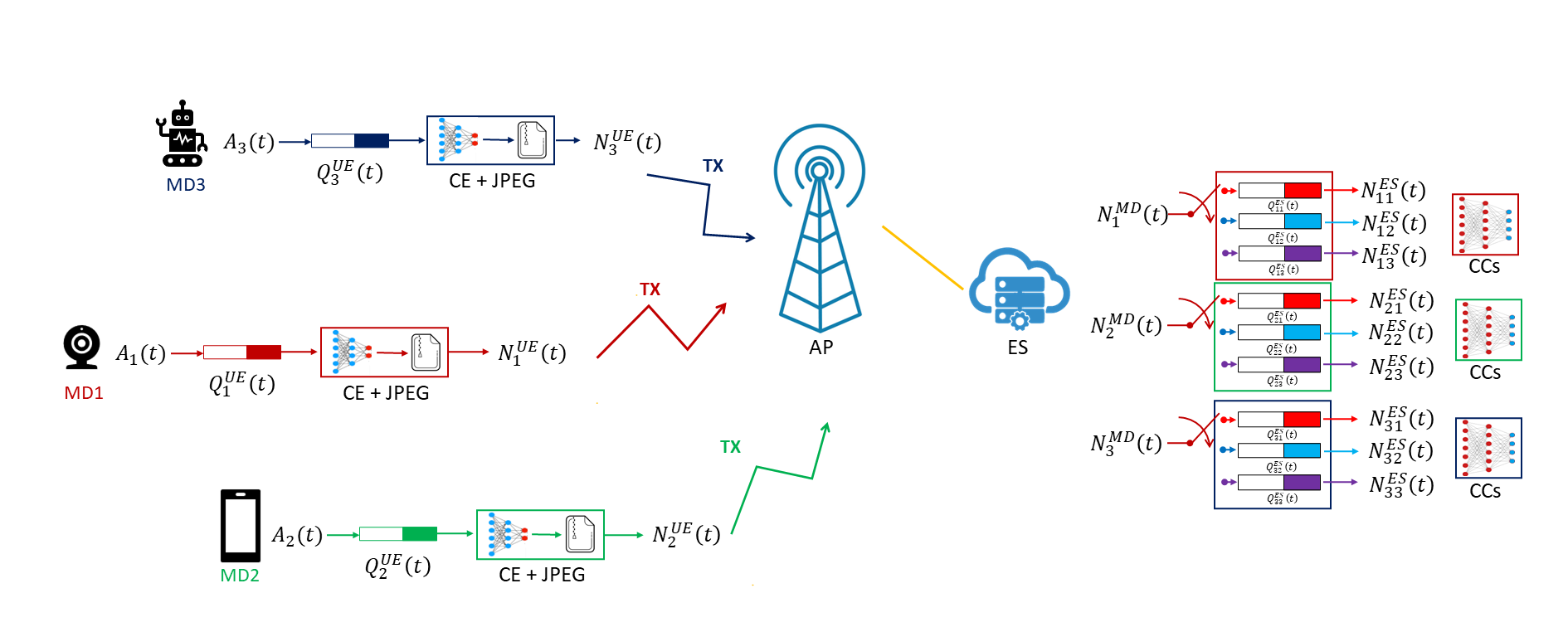}
    \caption{Scenario: each UE dinamically employs its own set of CEs coupled with a proper set of CCs at the ES.}
    \label{fig:system_scenario}
\end{figure*}

The considered goal-oriented scenario encompasses multiple devices (UEs), with limited computational and energy capabilities, which are connected through an Access Point (AP) to an ES with a larger amount of computing resources; an illustration is given in Fig. \ref{fig:system_scenario}. To perform a generic learning task, for each UE connected to the network, the system handles three main phases: i) The UE buffers the Data Units (DUs), i.e., the images to be classified; ii) Depending on the specific offloading decision, which is affected by the system status, the DUs are either scheduled to be compressed and transmitted by the goal-oriented compression strategy proposed in Section \ref{sec:training} or, alternatively, to be processed locally; iii) The inference task takes place either at the UE- or ES-side, depending on the offloading decision.

The system evolves in a time-slotted fashion, where each time slot has a fixed duration $\tau$. Therefore, we deal with discrete-time functions $f(t)$, where $t\in \mathbb{N}$ is an index for the $t$-th time-slot $[t\tau,(t+1)\tau[$.
The aim of the resource optimization strategies for GOC is to guarantee a specific E2E (maximum) delay requirement, while optimizing either the system energy consumption or the learning accuracy. To this end, the proposed policies have to manage several resources. In particular, the $k$-th UE has to allocate its \textit{transmission rate} $R_{k}(t)$ toward the ES, its \textit{clock frequency} $f_{k}^{d}(t)$, employed to perform the data compression by a specific \textit{compression factor} $\rho_{k}(t)$, and the offloading decision $d_{k}(t)$.
As far as the ES is concerned, the main optimization variable is represented by the \textit{clock frequency} $f_{c}(t)$, which has to be properly split among the learning tasks of the different users. This quantities represent the optimization variables of the objective functions we will define for the proposed resource management strategies. We are now ready to describe the models adopted for latency, energy and classification accuracy.

\subsection{Latency model}
The system evolution over time is entirely described by a queuing system, as prescribed by the Lyapunov optimization framework \cite{neely2010stochastic}. In particular, for each user involved in the network, we define two kind of physical queues:
\begin{itemize}
    \item[-] A \textit{computation/communication} queue at each UE, which collects the DUs, i.e., the images, generated by each device, which are waiting to be compressed and transmitted to the ES for classification.
    \item[-] A separate \textit{computation} queue at the ES side for any possible compression degree (e.g., CE) that the UEs may dynamically employ: thus, for each UE connected to the network, we have a different number of ES queues, depending on the CE compression degrees that are available. This design choice has been motivated in order to make the ES optimization problem computationally affordable, as we will clarify later.
\end{itemize}

We denote with $K$ the total number of UEs connected to the network. The binary variable $d_{k}(t)\in\{0,1\}$ models the decision to offload (or not) the learning task of the $k$-th device during the $t$-th time-slot. When any UE has to offload its learning task (i.e., $d_{k}(t)=1$), we make the following assumptions that are instrumental to practically manage the optimization problem (see  \cite{jasp_sumbitted_binucci} for further details).

\textit{Assumption 1: }The DUs in each UE queue have to be compressed and transmitted within the same time-slot. Indeed, during a given time-slot, it is impossible to optimally compress DUs that will be transmitted during one of the next time-slots, when the system could possibly experience different channel conditions, or different lengths of the ES/UEs queues, etc. Therefore, compression and transmission operations have to be done sequentially within the same time-slot.

\textit{Assumption 2: }We assume that, while an UE is transmitting some DUs, it can also simultaneously compress other DUs.

The number of (compressed) DUs that would be possible to transmit during the $t$-th time-slot is expressed by
\begin{equation}
	N_{k}^{tx}(t)=\floor*{\frac{\tau R_{k}(t)}{M(\rho_{k}(t))N(\rho_{k}(t))}},
\end{equation}
where $R_{k}(t)$ and $\rho_{k}(t)$ are the transmission rate and the compression factor{\footnote{Note that we denote with $\rho$ the compression factor of the images along each dimension. The actual compression ratio scales with $\rho^2$.}, respectively, selected for the $k$-th UE at time $t$;  $M(\rho_{k}(t))$ is the DU's size for a certain compression factor $\rho_{k}(t)$, and $N(\rho_{k}(t))$ is the number of bits that are necessary (on average) to encode a pixel in the (zipped) pseudo-image $h$. To shorten the notation, we define also $W(\rho_k(t))=M(\rho_k(t))N(\rho_k(t))$, which represents the average number of bits to store an image with a given $\rho_{k}(t)$. On the other hand, the number $N_{k}^{c}(t)$ of DUs that is possible to compress during the $t$-th time-slot by the $k$-th device is expressed by
\begin{equation}\label{device_compression_equation}
	N_{k}^{c}(t)=\floor*{\tau f_{k}^{d}(t)J_{d}(\rho_{k}(t))},
\end{equation}
where $J_{d}(\rho_{k}(t))$ denotes the number of DUs compressed in a clock cycle $C$ (which depends on the selected compression factor $\rho_{k}(t)$), and $f_{k}^{d}(t)$ denotes the device clock-frequency that has been chosen for the $k$-th UE, during the same time-slot. Recalling Assumption 1, all the DUs that are compressed within a time-slot have to be transmitted during the same time-slot, and all the transmitted DUs have to be first compressed. Thus, we need to use a transmission rate $R_{k}(t) \leq W(\rho(t))f_{k}^{d}(t)J_{d}(\rho(t))$ which results in $N_{k}^{tx}(t) \leq N_{k}^{c}(t)$. Taking into account that, before the transmission could start, we need to wait a time equal to $1/(f_{k}^{d}(t)J_{k}^{d}(t))$ to compress the first DU, the actual number of DUs that can be offloaded by the $k$-th device during the $t$-th slot is expressed by
\begin{equation}
	N_{k}^{off}(t)=\lf\frac{\tau-1/(f_{k}^{d}(t)J_{k}^{d}(t))}{W(\rho_{k}(t))/R_{k}(t)}\rf.
\label{eq:n_du_device}
\end{equation}
Plugging in \eqref{eq:n_du_device} the inequality $N_{k}^{tx}(t) \leq N_{k}^{c}(t)$ we end-up with the following (integer) inequality
\begin{equation}\label{N_UE_inequality}
\begin{aligned}
	\lf\frac{\tau R_{k}(t)}{W(\rho_{k}(t))}\rf-1 \leq N_k^{off}(t) \leq \lf \frac{\tau R_{k}(t)}{W(\rho_{k}(t))}\rf, 
	\end{aligned}
\end{equation}
which will be useful in the next derivations. 

Finally, similarly to (\ref{device_compression_equation}), when the learning task is performed locally, the total number of DUs processed by the $k$-th UE is expressed by
\begin{equation}
    N_{k}^{L}(t)=\floor{\tau
    f_{k}^{d}(t)J_{k}^{L}(\rho_{k}(t))},
\label{eq:n_class_device}
\end{equation}
where $J_{k}^{L}(\rho_{k}(t))$ expresses the DUs that can be compressed by a factor $\rho_{k}(t)$ and successively classified in a clock-cycle by the UE hardware. Putting together \eqref{eq:n_du_device} and \eqref{eq:n_class_device}, the number of DUs that can be processed by an UE, within a single time-slot, is expressed by
\begin{equation}
    N_{k}^{UE}(t)=d_{k}(t)\cdot N_{k}^{off}(t)+(1-d_{k}(t))\cdot N_{k}^{L}(t).
\label{eq:offloading_device_du}
\end{equation}
The UE queue $Q_{k}^{UE}(t)$ is fed by the arrival of new DUs, and is drained either by the transmission of DUs to the ES, or by their local classification at the UE. Thus, it is characterized by the following evolution
\begin{equation}
	Q_{k}^{UE}(t+1)=\max(0,Q_{k}^{UE}(t)-N_{k}^{UE}(t))+A_{k}(t),
\label{eq:device_queue}
\end{equation}
where $A_{k}(t)$ models the DUs arrival process, whose statistical properties are generally unknown.

At the ES, we employ $L_{k}$ different queues for each UE, whose evolution is described by
\begin{equation}
\begin{split}
    Q_{ki}^{ES}(t+1)&=\max(0,Q_{ki}^{ES}(t)-N_{ki}^{ES}(t))
    %breackequation \\ &
    \\
    &+d_{k}(t)\cdot \min(N_{k}^{UE}(t),Q_{k}^{UE}(t))\cdot\mathbbm{1}_{i}\{\rho_{k}(t)\},
\end{split}
\label{eq:server_queue}
\end{equation}
i.e., a queue for each compression factor among the $L_{k}$ in the set $\mathcal{S}_k=\{s_{ki}\}_{i=1,\ldots,L_k}$, which represents the set of the compression factors employable by the $k$-th UE.
These queues store the ES computation load, expressed in number of DUs, that is reserved for the $k$-th device.
The term $\mathbbm{1}_{i}\{\rho_{k}(t)\}$ in \eqref{eq:server_queue} is a shorthand for the indicator function $\mathbbm{1}\{\rho_{k}(t)=s_{ki}\}$, which models the arrival of new DUs in the ES queue only if the UE have chosen the $i$-th compression factor. 
The term $N_{ki}^{ES}(t)$ in \eqref{eq:server_queue} denotes the number of DUs processed by the ES during the $t$-th time-slot, and it is expressed by
\begin{equation}
    N_{ki}^{ES}(t)= \floor{\tau f_{ki}^{s}(t)J_{ki}^{s}(t)},
\label{eq:n_k_ser}
\end{equation}
where $f_{ki}^{s}(t)$ is the ES clock-frequency assigned to the $i$-th queue (compression factor) of the $k$-th UE, during the $t$-th time slot.\footnote{\label{Footnote-MultipleQueues} Having different queues for each compression factor is a design choice instrumental to obtain a mathematical dependence between $N_{ki}^{ES}$ and $f_{ki}^{s}$, that is simpler than in \cite{jasp_sumbitted_binucci}, where we used a single queue. This way, the solution of the ES optimization problem becomes feasible also in a multi-user context, as we will clarify later.}
The quantity $\frac{1}{J_{ki}^{s}(t)}$ in \eqref{eq:n_k_ser} is a conversion factor that maps the number of DUs received by the ES into the equivalent number of clock-cycles requested for their processing (e.g., classification).

To set-up our delay constraints, we need to define an overall queue that, for each device, takes into account the overall computational load at both the UE- and ES-side.
Since we aim to respect an average latency constraint, as we will detail in the following, and taking in mind the ES can perform a parallel computation of multiple DUs, by means of \eqref{eq:device_queue} and \eqref{eq:server_queue},
%we may consider an (overall) worse-case waiting queue associated to the $k$-th UE, as expressed (in terms of data-units) by
%\begin{equation}\label{qtot}
%	Q_{k}^{tot}(t)=Q_{k}^{UE}(t)+\max_{i}\{Q_{ki}^{ES}(t)\}.
%\end{equation}
it makes sense to consider the average length of the parallel queues, which is expressed by
\begin{equation}
	Q_{k}^{tot}(t)=Q_{k}^{UE}(t)+\sum_{i}^{L_{k}}p_{ki}Q_{ki}^{ES}(t),
\label{eq:total_queue}
\end{equation}
where $p_{ki}$ is the probability to employ the $i$-th compression factor in $\mathcal{S}_k$, which can be estimated by an online sample-mean\footnote{{The $p_{k,i}$ are actually time-varying with the system state, which is also influenced by the \emph{instantaneous} and adaptive resource management strategies we will end up with. The assumption here is that the stochastic resource management algorithms, which will exploit knowledge of the estimated $p_{k,i}$, will converge to a steady state where also the running sample mean estimate of the $p_{k,i}$ will converge. This fact has been verified by extensive simulation results.}}}. 
By assuming a certain data arrival rate $\overline{A_{k}}=\mathbb{E}\left\{\frac{A_{k}(t)}{\tau}\right\}$, and exploiting the Little's Law \cite{little1961proof}, \eqref{eq:total_queue} allow us to model the average long-term delay, as expressed by
\begin{equation}
	\lim_{T\to\infty} \frac{1}{T} \sum_{t=1}^{T} \mathbb{E}\left\{\frac{Q_{k}^{tot}(t)}{\overline{A_{k}}}\right\}.
\end{equation}

For a latency constraint $D_{k}^{avg}$, we get a queue length constraint $Q_{k}^{avg}=D_{k}^{avg}\overline{A}_{k}$ and, consequently, we can equivalently formalize the latency constraint as a queue constraint by
\begin{equation}
	\lim_{T\to\infty} \frac{1}{T} \sum_{t=1}^{T} \mathbb{E}\{Q_{k}^{tot}(t)\} \leq Q_{k}^{avg}.
\label{eq:queue_length_constraint}
\end{equation}

\subsection{Energy model}
The energy model of our system involves three main components:
\begin{itemize}
\item[-] \textit{Transmission energy at the UEs}, requested to transmit the DUs to the ES in case of offloading decisions.
\item[-] \textit{Computation energy at the UEs}, requested in order to either compress/encode the DUs to be transmitted, or to perform the learning task locally.
\item[-] \textit{Computation energy at the ES}, requested to classify the DUs transmitted by the UEs that decide to offload the learning tasks.
\end{itemize}
For simplicity, assuming a capacity achieving transmission system, in a flat-fading wireless channel, the transmission power $p_{k}^{tx}(t)$ requested by the $k$-th UE can be inferred by the Shannon capacity \cite{Shannon_Capacity}
\begin{equation}\label{eq:shannon_capacity}
R_{k}(t)=B_{k}\log_{2}\left(1+\frac{p_{k}^{tx}(t)|h_{k}(t)|^{2}}{N_{0}B_{k}}\right),
\end{equation}
where $|h_{k}(t)|$ is the channel gain, $N_{0}$ denotes the noise power spectral density at the receiver side, and $B_{k}$ is the bandwidth. Thus, by inverting \eqref{eq:shannon_capacity}, we obtain that the transmission energy spent by the $k$-th UE during the $t$-th time-slot depends on the rate $R_{k}(t)$ by
\begin{equation}
E_{tx}^{k}(t)=\tau p_{k}^{tx}(t)=\frac{\tau B_{k}N_{0}}{|h_{k}(t)|^{2}}\left(e^{\frac{R_{k}(t)ln(2)}{B_{k}}}-1\right).
\label{eq:transmission_energy}
\end{equation}
%Therefore, the transmission energy spent by the $k$-th UE during the $t$-th time-slot is expressed by
%\begin{equation}
%E_{tx}^{k}(t)=\tau p_{tx}^{k}(t).
%\label{eq:transmission_energy}
%\end{equation}
From the computation perspective, the ES's and UE's models are equivalent. Specifically, in order to estimate the energy consumption, we exploit the model in \cite{processor_energy_article}, which assumes a cubic dependence on the ES's and UE's clock-frequencies $f_s(t)$ and $f_{k}^{d}(t)$, as expressed by
\begin{equation}
	 E_{k}^{d}(t)  = \tau\kappa_{k}^{d}  f_{k}^{d}(t)^{3} \hspace{12pt} \hbox{and} \hspace{12pt}
  E_{s}(t)  = \tau\kappa_s  f_{s}(t)^{3}.
\label{eq:comp_energy}
\end{equation}
The constants $\kappa_s$ and $\kappa_k^d$ represent the effective switched capacitance \cite{processor_energy_article} of ES and  $k$-th UE processor, respectively. %For simplicity we assume the same $\kappa$ for all the devices and the server, although this is not strictly necessary.
Thus, we quantify the system energy consumption during the $t$-th time-slot using the following weighted performance metric:
\begin{equation}
E_{k}^{tot}(t)=(1-\gamma)E_{s}(t)+\gamma \sum_{k=1}^{K}\delta_{k}(E_{k}^{c}(t)+E_{k}^{tx}(t)),
\end{equation}
where the parameter $\gamma$ is used to weight the UEs versus ES energy consumption, enabling tuning toward the implemention of an user-centric $(\gamma \to 1)$ or a server-centric $(\gamma \to 0)$ optimization strategy. Furthermore, the weights $\{\delta_{k}\}_{k=1}^K$ (with $\sum_{k=1}^{K}{\delta_k}=1$) can be employed to assign different importance to the the energy consumption of different users, providing an extra degree of flexibility to the optimization, depending on the needs of the operators, users, and service providers.

\subsection{Accuracy model}
\label{sec:Accuracy_Model}
For the accuracy of the learning task of each UE, we resort to a {\it model-based} management strategy. This means that the accuracy for the $k$-th task can be cast in the optimization problem as a function $G_k(\rho_{k}(t))$ of the compression degree. This can be done in practice by employing a look-up table (LUT) (shown in sec.\ref{sec:sim_res}), where each entry is associated with a specific compression factor $\rho_{k} \in \mathcal{S}_k$.\footnote{We modeled the relationship between the compression factor and the accuracy through a LUT, rather than by a formal analytical expression, because it is   almost impossible to find a closed-form expression for this function in practice. Indeed, despite noticeable examples to theoretically formalize DNNs performance can be found in \cite{tishby2015deep},\cite{saxe2019information}, these approaches are based on Mutual Information, which is intractable to derive in closed-form in most of the practical cases.}
This LUT stores the (average) classification accuracy of the $k$-th learning task, associated with each one of the CE-CC classifying chains that are available for the $k$-th UE. The values stored in this accuracy-LUT can be estimated off-line on meaningful test-sets, after each CE-CC structure has been properly trained, as described in the previous section. Thus, we can exploit the LUTs $G(\rho_{k}(t))$ to enforce an average accuracy constraint for each learning task, as expressed by
\begin{equation}
	\lim_{T\to\infty} \frac{1}{T} \sum_{t=1}^{T} \mathbb{E}\{G_{k}(\rho_{k}(t))\} \geq G_{k}^{avg}.
\end{equation}

\section{Dynamic Resource Optimization for Multi-user Goal-oriented Communications}\label{sec:opt_strat_develop}
On the basis of the delay, accuracy, and energy models presented in the previous section, we develop two resource optimization strategies: a multi-user Minimum-Energy with (maximum) Delay and Accuracy constraints (mu-MEDA), and a multi-user Maximum-Accuracy with (maximum) Delay and Energy consumption constraints (mu-MADE). In the sequel, we describe the problem formulation and the algorithmic solution for both strategies.
\subsection{mu-MEDA: multi-user Minimum-Energy with Delay and Accuracy constrains}
\label{subsec:mu-MEDA}
Following a system energy minimization perspective, the long-term optimization problem can be cast as follows:
\begin{mini}|s|[2]
{\Phi(t)}{\lim_{T\to\infty} \frac{1}{T} \sum_{t=1}^{T} \mathbb{E}\{E_{tot}(t)\}}{\label{eq:long_term_meras}}{}
\addConstraint{(a)\:}{\lim_{T\to\infty} \frac{1}{T} \sum_{t=1}^{T} \mathbb{E}\{Q_{k}^{tot}(t)\}}~{\leq Q_{k}^{avg},\forall k}
\addConstraint{(b)\:}{\lim_{T\to\infty} \frac{1}{T} \sum_{t=1}^{T} \mathbb{E}\{G(\rho_{k}(t)\}\geq G_{k}^{avg}, \forall k}{} 
\addConstraint{(c)\:}{\; 0 \leq R_{k}(t) \leq R_{k,max}, \quad \forall k,t} 
\addConstraint{(d)\:}{\;\rho_{k}(t) \in \mathcal{S}_{k},\; \quad f_{s}(t) \in \mathcal{F}_{s}, 
%breackequation}{
\; \quad f_{k}^{d}(t) \in \mathcal{F}_{d,k}
%breackequation}{
\quad \forall k,t}
\addConstraint{(e)\:}{\sum_{k=1}^{K}\sum_{i=1}^{L_{k}}f_{ki}^{s}(t) \leq f_{s}(t),
\quad (f)\:
%breackequation }{
f_{ki}^{s}(t) \geq 0}
\addConstraint{(g)\:}{d_{k}(t) \in \{0,1\}\; \quad \forall k,i,t}{\;}
\end{mini}
\vspace{0.5pt}

where $\Phi(t)=[\{R_{k}(t),f_{ki}^{s}(t),f_{k}^{d}(t), \rho_{k}(t), d_{k}(t)\}_{k=1}^K,f_{s}(t)]$ contains all the optimization variables. The constraints in \eqref{eq:long_term_meras} have the following meaning: $(a)$ the average queue length for the $k$-th UE must be lower than $Q^{avg}_{k}$, i.e., we are imposing a maximum average service delay equal to $D^{k}_{avg}=Q^{k}_{avg}/\overline{A_{k}}$ (cf. \eqref{eq:queue_length_constraint}); $(b)$ the average classification accuracy for the $k$-th UE must be greater that $G^{k}_{avg}$; $(c)$ the $k$-th UE transmission rate $R_{k}(t)$ must be smaller than the value $R_{k,max}(t)$, which is the maximum possible rate for the $k$-th device, inferred by \eqref{eq:shannon_capacity}, considering the maximum available transmission power $p^{tx}_{k,max}$; $(d)$ specifies the discrete sets $\mathcal{F}_{c}$, $\mathcal{F}_{d,k}$ and $\mathcal{S}_{k}$ for the server frequencies set, the frequencies set for the $k$-th UE, and the set  of the possible compression factors respectively; the constraints $(e)-(f)$ state that the sum of the clock frequencies $f_{ki}^s(t)$ that the (edge) server allocates for all the queues assigned to each user, must be lower than the total ES clock-frequency chosen for the $t$-th time slot, and that each clock-frequency must be obviously grater than 0; finally, $(g)$ represents the binary constraints on the set of the opportunistic offloading decisions variables of each UE. Problem (\ref{eq:long_term_meras}) is complicated due to the lack of knowledge of the statistics of the radio channels and data arrivals, which would be necessary to compute the expected values in \eqref{eq:long_term_meras}. To tackle this issue, we resort to Lyapunov stochastic optimization arguments \cite{neely2010stochastic}, which solve the long term problem \eqref{eq:long_term_meras} by casting it to a sequence of instantaneous optimization problems, which can be solved in a per-slot fashion. According to such an optimization framework \cite{neely2010stochastic}, we start associating a {\it virtual} queue to each one of the long-term constraints $(a)$ and $(b)$. These virtual queues evolve according to
\begin{equation}
\begin{split}
    Z_{k}(t+1) &= \max(0,Z_{k}(t)+\mu_{k}(Q_{k}^{tot}(t+1)-Q_{k}^{avg}))
    \\
    Y_{k}(t+1) &= \max(0,Y_{k}(t)+\nu_{k}(G_{k}^{avg}-G_{k}(t))),
\end{split}
\label{eq:mu_meras_virtual_queues}
\end{equation}
where $\mu_{k}$ and $\nu_{k}$ are step-sizes that control the convergence speed of the algorithm. This way, it is possible to prove that respecting the long term constraints $(a)-(b)$ is equivalent to  guarantee the mean-rate stability of the virtual queues in \eqref{eq:mu_meras_virtual_queues} \cite{neely2010stochastic}. To this end, we define the Lyapunov function $L(t)$, as the sum of the squares of all the (virtual and physical) queues
\begin{equation}
    L(t)=\sum_{k=1}^{K}Z_{k}(t)^{2}+\sum_{k=1}^{K} Y_{k}(t)^{2}.
\end{equation}
Defining $\Theta(t) = \left[ \{Z_{k}(t)\}_{k=1}^{K},\{Y_{k}(t)\}_{k=1}^{K}\right]$, we obtain the associated conditional \textit{Lyapunov drift}
\begin{equation}
    \Delta(\Theta(t))=\mathbb{E}\{L(t+1)-L(t)|\Theta(t)\},
\label{eq:lyapunov_drift}
\end{equation}
whose minimization corresponds to the stabilization of the virtual queues, but it does note take into account the objective function (i.e., the system energy consumption). Thus, in order to trade-off system stability and energy consumption, the Lyapunov Drift is augmented with a term dependent on the system energy, to obtain the so-called \textit{Lyapunov Drift plus Penalty} function
\begin{equation}
    \Delta_{p}(\Theta(t))=\Delta(\Theta(t))+V\mathbb{E}\{E_{tot}(t)\}.
\label{eq:lyapunov_drift_plus_pentalty}
\end{equation}
By increasing the value of the parameter $V$ we give more importance to the objective function rather than to the queues stability, thus pushing the solution toward optimality while still guaranteeing the stability of the system, i.e., respecting the long-term constraints. In particular,  \cite{neely2010stochastic} proved that, as the parameter $V$ increases, the optimal solution of \eqref{eq:long_term_meras} is asymptotically reached. Following stochastic optimization arguments \cite{neely2010stochastic}, we proceed minimizing an upper bound of the Lyapunov Drift plus penalty function in \eqref{eq:lyapunov_drift_plus_pentalty} (derived in the Appendix), ending up with the instantaneous optimization problem in \eqref{eq:overall_optim_meras_problem_1}, where,  since the optimization variables affect only the terms $N_{k}^{UE}$, $N_{ki}^{ES}$ and $G_{k}$, we neglect all the terms which do not depend on them. Note moreover that in the following we omit the time index $t$ to simplify the notation.
\begin{align}
\label{eq:overall_optim_meras_problem_1}
&\!\min_{\Phi}   && VE_{tot}+\sum_{k=1}^{K} \bigg[L_{k}N_{k}^{UE}\mu_{k}^{2}\left(\sum_{i=1}^{L_{k}}\mathbbm{1}_{i}\{\rho_{k}\}p_{ki}Q_{ki}^{ES} -Q_{k}^{UE}\right)\notag \\ 
%breackequation \notag \\ & &&
& && -L_{k}\mu_{k}^{2}\sum_{i=1}^{L_{k}}p_{ki}Q_{ki}^{ES}N_{ki}^{ES} +\mu_{k}Z_{k}(\max(0,Q_{k}^{UE}-N_{k}^{UE})
\notag \\ 
&  &&
%breackequation \notag\\ &  &&
+\sum_{i=1}^{L_{k}}\max(0,p_{ki}Q_{ki}^{ES}-N_{ki}^{ES})) -\nu_{k}Y_{k}G_{k}(\rho_{k}) \bigg] 
%breackequation \notag\\ &  &&
\notag
\\
%breackequation &   &&\notag\\
&\!\text{s.t.} && 0 \leq R_{k} \leq R_{k,max}, \quad 
%breackequation \qquad \qquad \forall k,t \notag\\ &    && 
\rho_{k} \in \mathcal{S}_{k}, 
%breackequation \hspace{3pt} 
\quad f_{s} \in \mathcal{F}_{s}, 
%breackequation \hspace{3pt} 
\quad f_{k}^{d} \in \mathcal{F}_{d,k} 
\\ \notag
%breackequation \qquad\forall k,t \notag\\ & &&
& && \sum_{k=1}^{K}f_{k}^{s} \leq f_{s,} 
%breackequation \qquad \qquad \qquad \forall t \notag\\ &  &&
\quad f_{k}^{s} \geq 0, \quad
%breackequation \quad\qquad \qquad \qquad \qquad \qquad \quad 
\forall k,t. \notag
\end{align}

Since the UEs energy-consumption terms in the cost function of problem \eqref{eq:overall_optim_meras_problem_1} depend only (and separately for each UE) on the UEs optimization variables $\{\Phi_{d,k}\}_{k=1}^{K}=\{[R_{k},f_{k}^{d},\rho_{k},d_{k}]\}_{k=1}^{K}$, we can optimize this part of the cost function separately at each UE. Note that our design choice to assign at the ES separate computation queues for each UE offloaded task, lets us completely decouple the optimization problem and separately handle the UE and ES resource optimization.
%, as detailed in the following. 
Furthermore, as already pointed out in footnote \ref{Footnote-MultipleQueues}, the use of multiple queues for each compression factor $\rho_{ki}$, thanks to \eqref{eq:total_queue}, makes by \eqref{eq:n_k_ser} the problem linear with respect to $f_{ki}$, up to the $\lfloor . \rfloor$ operator. Consequently, Problem \eqref{eq:overall_optim_meras_problem_1} is separable and solvable for each compression factor, as described in the following.

\subsubsection{UE sub-problem}
For the $k$-th device, at each time slot $t$, we have to solve the following optimization problem
\begin{align}
\min_{\Phi_{d,k}}  \quad & L_{k}N_{k}^{UE}\mu_{k}^{2}\bigg(\sum_{i=1}^{L_{k}}\mathbbm{1}_{i}\{\rho_{k}\}p_{ki}Q_{ki}^{ES}-Q_{k}^{UE}\bigg) \notag  \\ 
%breackequation \notag \\ & 
& +\mu_{k}Z_{k}\max(0,Q_{k}^{UE}-N_{k}^{UE}) 
-\nu_{k}Y_{k}G_{k}(\rho_{k})\notag \\
&+V\gamma\delta_{k}(E_{k}^{tx}+E_{k}^{c})
\\[0.5pt]
    \text{s.t.}\quad &  0 \leq R_{k} \leq R_{k,max}
    %breackequation \\&
    ,\quad \rho_{k} \in \mathcal{S}_{k},\quad f_{k}^{d}(t) \in \mathcal{F}_{d,k},\\
    \quad &  d_{k} \in \{0,1\} \notag.
           \end{align}
%\end{equation}
% where $\Phi_{d,k}=[R_{k},f_{k}^{d},\rho_k,d_{k}]$.
Depending on the value of the offloading decision variable $d_k$ we can optimize the other variables employing two different strategies. If $d_{k}=1$, we have to allocate both the transmission rate $R_k$ to transmit the DUs to the ES, and the UE clock-frequency $f_{k}^{d}$ and compression factor $\rho_k$ to perform compression. Otherwise, if $d_{k}=0$ we need only to allocate $f_{k}^{d}$ and $\rho_k$ to perform the learning task locally. We remark that we assume, although this is not mandatory, that the UE employs also locally the same (bank of) CE-CC classification chains we designed for the GOC scheme, thus fairly offering to the UEs the same flexibility of classification accuracy and energy consumption that could be exploited by the ES solution. Other choices, or a fixed structure of the classifier at the UE, would obviously have an impact on the offloading decisions by the optimal resource management and, consequently, on the energy-delay-accuracy tradeoffs.

Coming to the solution of the problem, when $d_{k}=1$ we handle the $\min(\cdot)$ in \eqref{eq:n_du_device} by adding the following constraint on the transmission rate of the $k$-th user
\begin{equation}
\begin{split}
    & 0 \leq R_{k}(t) \leq R_{k,max}^{+}(t),
    \quad\\
    & R_{k,max}^{+}=\min\bigg\{R_{k,max},\frac{\small{Q_{k}^{UE}W(\rho_{k})}}{\small{\tau}}
    %,J_{d}(\rho_{k})f_{k}^{d}W(\rho_{k}) 
    \bigg\}.
\end{split}
\end{equation}
This way, according to Assumptions 1 and 2, and taking in mind we cannot compress more DUs that we can transmit, we select a data-rate that is bounded by the minimum between the {\it maximum achievable rate} $R_{k,max}$ (computed plugging the maximum power $p_k^{tx}$ in the Shannon capacity \eqref{eq:shannon_capacity}), and the {\it draining rate} $Q_{k}^{UE}W(\rho_{k})/\tau$ that is capable to empty the transmission queue (and lets remove the $\max(\cdot)$).
% and the {\it compression rate} that lets to transmit all the DUs that is possible to compress in a time-slot.
By considering that $x-1 \leq \floor{x} \leq x$, we can also remove the $\floor{\cdot}$ in  \eqref{eq:n_du_device}. Therefore, using the definition of the indicator function,
for any {\it fixed} compression factor $\rho_{ki} \in \mathcal{S}_k$,
we end up with the following optimization problem
\begin{align}
\label{eq:device_per_cf_problem}
%\begin{alignat}{2}
\min_{\Phi_{d,k}}   &\quad- \frac{Q_{ki}^{TX}\tau R_{k}}{W(\rho_{k})} +  \frac{\tau V\gamma\delta_{k} B_{k}N_{0}}{h_{k}^{2}}e^{\frac{R_{k}ln(2)}{B_{k}}} 
%breackequation \notag \\ &\qquad 
+\tau V\gamma\delta_{k}\kappa (f_{k}^{d})^{3} \notag \\
& \quad -\nu_{k}Y_{k}G_{k}(\rho_{k})\notag\\[0.5pt]
\text{s.t.} &\quad  0 \leq R_{k} \leq R_{k,max}^{+}, %\tag{\ref{eq:device_per_cf_problem}}
\quad
%\rho_{k} \in \mathcal{S}_{k}, \quad
f_{k}^{d} \in \mathcal{F}_{d,k}, \notag %\end{alignat}
%\end{subequations}
\end{align}
\noindent
where $Q_{ki}^{TX}=L_{k}\mu_{k}^2(Q_{k}^{UE}-p_{ki}Q_{ki}^{ES})+\mu_{k}Z_{k}$.
This is a mixed-integer optimization problem. However, in practice, the sets $\mathcal{F}_{d,k}$ and $S_{k}$ have a quite low cardinality and, as detailed below, the solution can be rapidly found by an exhaustive search.
%Now, considering the optimization problem related to the $i$-th compression factor, fixing the device clock frequency, we have a convex optimization problem with respect to the data rate, whose solution can be found considering \emph{duality theory}
Indeed, for any fixed couple of compression factor $\rho_{k} \in \mathcal{S}_k$ and computation frequency $f_k^d \in \mathcal{F}_{d,k}$, the optimization problem is convex with respect to the data rate $R_k$, whose optimal value can be found in closed form by \emph{duality theory} through the Lagrangian
\begin{equation}
\begin{split}
\mathcal{L}=&-\frac{\tau Q_{ki}^{TX} R_{k}}{M(\rho_{k})N(\rho_{k})}+\frac{\tau V\gamma\delta_{k}N_{0}B_{k}}{h_{k}^{2}}e^\frac{R_{k}ln(2)}{B_{k}}
%breackequation \\&\quad 
+\tau V\gamma\delta_{k}\kappa (f_{k}^{d})^3\\
&-\nu_{k} Y_{k}G_{k}(\rho_{k})-\alpha R_{k}
%breackequation \\ &\qquad 
+ \beta(R_{k}-R^{+}_{k,max}),
\end{split}
\label{eq:optim_device}
\end{equation}

where $\alpha$ and $\beta$ are the Lagrangian multipliers. Note that, if $Q_{ki}^{TX} \leq 0$, the second term monotonically increases with the rate, and the minimum of the Lagrangian is obtained for $R_{k}=0$. Otherwise, when $Q_{ki}^{TX} > 0$ we can solve the optimization problem by imposing the following KKT conditions \cite{boyd2004convex}
\vspace{0.5pt}

\begin{equation}
\begin{split}
&(a) \;\frac{\partial \mathcal{L}}{\partial R_{k}} = -\frac{Q_{ki}^{TX} \tau}{W(\rho_{k})}+\frac{\tau V\gamma\delta_{k}ln(2)N_{0}B_{k}}{h_{k}^{2}}e^\frac{R_{k}ln(2)}{B_{k}}\\
&-\alpha+\beta=0\\
&(b) \; 0 \leq R_{k} \leq R^{+}_{k,max}
,\; (c) \; \alpha \geq 0
,\; (d) \; \beta \geq 0 \\ 
& (e) \; \alpha R_{k}=0
,\; (f) \; \beta (R_{k}-	R^{+}_{k,max})=0.
\end{split}
\end{equation}

Solving the KKT conditions leads to the following equation to compute the optimal rate
\begin{equation}\label{R_star}
R_{k}^{*}(\rho_{k},f_{k}^{d})=
\begin{cases}
    \left[\frac{B_{k}}{\ln(2)}\ln\left(\frac{Q_{ki}^{TX}h_{k}^{2}}{W(\rho_{k})V\gamma\delta_{k}\ln(2)N_{0}}\right)\right]^{R^{+}_{max}}_{0} &\text{\hspace{-.5em}\scriptsize{$Q_{ki}^{TX} > 0$}}\\
0 &\text{\hspace{-.5em}\scriptsize{otherwise}}
\end{cases}
\end{equation}
% \small
% $R_{k}^{*}(\rho_{k},f_{k}^{d})$
% \begin{equation}\label{R_star}
% R_{k}^*(\rho_k,f_{k}^{d}) = \begin{cases}
% \displaystyle \left[\frac{B_{k}}{\ln(2)}\ln\left(\frac{Q_{ki}^{TX}h_{k}^{2}}{W(\rho_{k})V\gamma\delta_{k}\ln(2)N_{0}}\right)\right]^{R^{+}_{max}}_{0} &\text{\hspace{-.5em}, if $Q_{ki}^{TX} > 0$;}
% \\
% 0 &\text{\hspace{-.5em}, otherwise.}
% \end{cases}
% \end{equation}
% \normalsize
which gives us the closed form expression for the optimal rate for any fixed compression factor $\rho_{k}$ and clock frequency $f_k^d$, of the $k$-th user. Thus, as anticipated, to select the best clock frequency $f_{k}^{d*}$, and compression factor $\rho_k^*$, we can proceed by an exhaustive search,  thanks to the limited cardinality of $\mathcal{F}_{d,k}$ and $\mathcal{S}_k$.
Summarising, for a potential offloading ($d_k=1$), we compute the optimal rate and clock frequency $f_k^d$ for each possible compression factor $\rho_{k}$, and then, at every time slot, we select the triple $T_{k}^{*}=(R_{k}^{*},f_{k}^{d*},\rho_{k}^{*})$ that gives the lowest energy cost.
%, for every time-slot. 
Otherwise, for a potential classification at the UE ($d_{k}=0$), the transmission rate to the ES would be $R_{k}=0$ and we need to optimize only the clock-frequency for each possible compression factor, thus obtaining the optimal pair $P_{k}^{*}=(f_{k}^{d*},\rho_{k}^{*})$ that minimizes the UE's energy consumption.
The overall optimal solution of the UE's optimization problem, which includes the decision to offload or not the learning task, is finally given by choosing between the pairs $(d_{k}=1,T_{k}^{*})$ and  $(d_{k}=0,P_{k}^{*})$, as the one that leads to the minimum value of the UE's energy cost function. 
%The overall procedure for UE resource allocation is summarized in Algorithm 1.
\subsubsection{ES sub-problem}
From the ES perspective, for each UE we have to manage multiple computing queues, each one associated to a specific compression factor that has been used by the specific UE: in the following, we denote with $Q_{ki}^{ES}$ the $i$-th ES computing queue for the $k$-th UE. It clearly  makes sense to constrain the fraction $f_{ki}^s$ (of the the total ES's computing frequency $f_s$) reserved to the $i$-th queue of the $k$-th user, to be lower than what would be necessary to completely drain the same queue within a time-slot, as expressed by
\begin{equation}
    f_{ki}^{s}(t) \leq \min\left(f_{s}(t),\frac{Q_{ki}^{ES}(t)}{\tau J_{ki}^{s}(t)}\right).
\end{equation}
This way, we can remove the terms $\max(0,Q_{ki}^{ES}-N_{ki}^{ES})$ from the sum in \eqref{eq:overall_optim_meras_problem_1} and, consequently, we can rewrite the ES's resource allocation problem as
\begin{align}
\label{eq:server_prob}
\min_{\Phi_{s}}  &\quad -\sum_{k=1}^{K} \sum_{i=1}^{L_{k}} \tau Q_{ki}^{comp} J_{ki}^{s}f_{ki}^{s}+\tau V(1-\gamma)\kappa f_{s}^{3}
\\[0.5pt]
\text{s.t.} &\quad  0 \leq f_{ki}^{s}(t) \leq \min\left(f_{s},\frac{Q_{ki}^{ES}}{\tau J_{ki}^{s}}\right) %\forall k,t;
%breackequation \notag\\&
\notag \\
& \quad \sum_{k=1}^{K}\sum_{i=1}^{L}f_{ki}^{s} \leq f_{s}, \quad f_{s} \in \mathcal{F}_{s}, \notag
\end{align}

%\end{subequations}
where $\Phi_{s}=[\{f_{ki}^{s}\}_{i=1,\ldots,L_k, k=1,\ldots,K}, f_{s}]$, and $Q_{ki}^{comp}=L_{k}\mu_{k}^2Q_{ki}^{ES}+\mu_{k}Z_{k}$. Although the problem is a mixed-integer optimization one, for any fixed ES's clock frequency $f_{s}$, it boils down to the classical (fractional) knapsack problem \cite{Knapsack_Problem}. Consequently, the optimal solution is obtained by a greedy algorithm, which consists in ordering the queues by their weights ($Q_{ki}^{comp}J_{ki}^{s}$) in descending order, and then assigning the clock frequency to the queue as $\min\left(\overline{\phi},\frac{Q_{ki}^{ES}}{\tau J_{ki}^{s}}\right)$, where $\overline{\phi}$ is the remaining part of the ES's clock frequency $f_{c}(t)$. Due to the limited cardinality of the ES's clock-frequency set $\mathcal{F}_s$, also in this case we can exhaustively solve the problem for all the server clock frequencies $f_{s} \in \mathcal{F}_s$, thus obtaining the set of possible solutions $\{(f_{ki}^{s},f_s )\}_{f_s\in\mathcal{F}_s}$ and then choose the one associated with the minimum ES's cost in \eqref{eq:server_prob}.

\subsection{mu-MADE: multi-user Maximum-Accuracy with Delay and Energy constraints}\label{subsec:mu-MADE}

An alternative resource allocation, targeting a Maximum-Accuracy, can be formulated as 
\begin{mini}|s|[2]
{\Phi(t)}{\lim_{T\to\infty} \frac{1}{T} \sum_{t=1}^{T} \mathbb{E}\left\{\sum_{k=1}^{K}-G_{k}(t)\right\}}{}{}
\addConstraint{\quad (a)\:}{\lim_{T\to\infty} \frac{1}{T} \sum_{t=1}^{T} \mathbb{E}\{Q_{k}^{tot}(t)\}\leq Q_{k}^{avg}}
\addConstraint{\quad (b)\:}{\lim_{T\to\infty} \frac{1}{T} \sum_{t=1}^{T} \mathbb{E}\{E_{k}^{d}(t)\}\leq E_{k}^{d,avg}}{\; \quad \forall k}
\addConstraint{\quad (c)\:}{\lim_{T\to\infty} \frac{1}{T} \sum_{t=1}^{T} \mathbb{E}\{E_{s}(t)\}\leq E_{s}^{avg}}
\addConstraint{\quad (d)\:}{0 \leq R_{k}(t) \leq R_{k,max} }{\;\quad  \forall k,t}
\addConstraint{\quad (e)\;}{\rho_{k}(t) \in \mathcal{S}_{k}, \; f_{s}(t) \in \mathcal{F}_{s},\;f_{k}^{d}(t) \in \mathcal{F}_{d,k}}{\; \quad \forall k,t}
\addConstraint{\quad (f)\:}{\sum_{k=1}^{K}\sum_{i=1}^{L_{k}}f_{ki}^{s}(t) \leq f_{s}(t)}{\; \quad \forall k,t}
\addConstraint{\quad (g)\:}{f_{ki}^{s}(t) \geq 0}{\; \quad \forall k,i,t}
\addConstraint{\quad (h)\:}{d_{k}(t) \in \{0,1\}}{\; \quad \forall k}
\end{mini}
\vspace{1pt}

where $\Phi(t)=[\{R_{k}(t),f_{ki}^{s}(t),f_{k}^{d}(t), \rho_{k}(t), d_{k}(t)\},f_{s}(t)]$, for $k=1,\dots,K$, and $i=1,\dots,L_k$  contains all the optimization variables. The constraints in \eqref{eq:long_term_meras} have the following meaning: $(a)$ the average queue length for the $k$-th UE must be lower than $Q^{avg}_{k}$, i.e., we are imposing a maximum average service delay equal to $D^{k}_{avg}=Q^{k}_{avg}/\overline{A_{k}}$ (cf. \eqref{eq:queue_length_constraint}); $(b)$ the average energy consumption for the $k$-th UE must be lower than $E^{k}_{d,avg}$; $(c)$ the average ES's energy consumption must be lower than $E^{s}_{avg}$; $(d)$-$(h)$ are equivalent to $(c)$-$(g)$ in \eqref{eq:long_term_meras}.

Proceeding similarly to the mu-MEDA strategy, in order to manage the long-term energy constraints (b) and (c), in addition to the virtual queue $Z_{k}(t)$ defined in \eqref{eq:mu_meras_virtual_queues} to manage $(a)$, we need to define the virtual queues
\begin{equation}
\begin{split}
    S_{k}(t+1)&=\max(0,S_{k}(t)+\lambda_{k}(E_{k}^{d}(t+1)-E_{k}^{d,avg}))\\
    O(t+1)&=\max(0,O_{k}(t)+\eta(O(t+1)-E_{s}^{avg})),
\end{split}
\label{eq:mu_maras_virtual_queues}
\end{equation}
where $\{\lambda_{k}\}_{k=1}^{K}$ and $\eta$ are the step-sizes used to control the convergence speed of the algorithm.
By the definition of the virtual queues, in this case the Lyapunov Function becomes
\begin{equation}\label{eq:lyapunov_function_maras}
    L(t)=\sum_{k=1}^{K}[S_{k}(t)^2+Z_{k}(t)^2]+O(t)^2
\end{equation}
and, consequently, given $\Theta(t)=[\{S_{k}(t),Z_{k}(t)\}_{k=1}^{K},O(t)]$, we derive the following expression for the Lyapunov \textit{drift-plus-penalty} function
\begin{equation}
    \Delta_{p}(t)=\mathbb{E}\{L(t+1)-L(t)|\Theta(t)\}-V\mathbb{E}\left\{\sum_{k=1}^{K}G_{k}(t)\right\}
\label{eq:mu_meras_lyap_drift_plus_penalty}
\end{equation}
As detailed in the Appendix, we end up with the following optimization problem

\begin{align}
\hspace{0cm} \min_{\Phi}  & \sum_{k=1}^{K} \bigg[L_{k}N_{k}^{UE}\mu_{k}^{2}\left(\sum_{i=1}^{L_{k}}\mathbbm{1}_{i}\{\rho_{k}\}p_{ki}Q_{ki}^{ES} -Q_{k}^{UE}\right) \notag \\
&+\mu_{k}Z_{k}(\max(0,Q_{k}^{UE}-N_{k}^{UE})+\sum_{i=1}^{L_{k}}\max(0,p_{ki}Q_{ki}^{ES}-N_{ki}^{ES})) \notag \\
&+\lambda_{k}S_{k}E_{k}^{d} -L_{k}\mu_{k}^{2}\sum_{i=1}^{L_{k}}p_{ki}Q_{ki}^{ES}N_{ki}^{ES}\bigg] +\eta OE_{k}^{s} -V\sum_{k=1}^{K}G_{k}(t)\notag
%breackequation \notag \\ &\quad 
% \notag
\\[0.5pt]
\hspace{0cm}\text{s.t.} &\quad 0 \leq R_{k} \leq R_{k,max},
%breackequation \qquad \forall k,t \notag\\ & 
\; \rho_{k} \in \mathcal{S}_{k},\; f_{s} \in \mathcal{F}_{s},\; f_{k}^{d} \in \mathcal{F}_{d,k} \notag \\
%breackequation \quad\forall k,t \notag\\ & 
& \sum_{k=1}^{K}f_{k}^{s}(t) \leq f_{s}(t),\; f_{k}^{s} \geq 0, \quad \forall k,t. 
%\end{alignat}
\label{eq:overall_optim_meras_problem_2}
\end{align}

Exploiting again the decoupling of the problem, which is granted by our proposed design to separately handle the queues for any specific UE and any specific compression factor, we end-up also in this case with distinct instantaneous optimization problems, one at each UE, and a single one at the ES.

\subsubsection{UE sub-problem}
As far as the $k$-th UE is concerned, we get the following optimization problem formulation
\begin{equation}\label{eq:mu_Maras_UE_cost_function}
\begin{aligned}
\min_{\Phi_{d,k}}       &\quad L_{k}N_{k}^{UE}\mu_{k}^{2}\left(\sum_{i=1}^{L_{k}}\{\mathbbm{1}_{i}\{\rho_{k}\}p_{ki}Q_{ki}^{ES}\-Q_{k}^{UE}\right) \\
& +\mu_{k}Z_{k}\max(0,Q_{k}^{UE} \- N_{k}^{UE})
\\
& -\lambda_{k}S_{k}E_{k}^{d}-VG_{k}(\rho_{k})\\
\text{s.t.} &\quad  0 \leq R_{k}(t) \leq R_{k,max}\\ 
%breackequation \qquad\forall k,t\notag \\ &
& \quad \rho_{k}(t) \in \mathcal{S}_{k},\hspace{3pt} f_{k}^{d}(t) \in \mathcal{F}_{d,k},\hspace{3pt} d_{k} \in \{0,1\},\hspace{6pt} \forall k,t 
\end{aligned}
\end{equation}
%\end{alignat}
%\end{subequations}
where $\Phi_{d,k}=[R_{k},f_{k}^{d},\rho_k,d_{k}]$, for $k=1,\dots,K$.
%\normalsize
The resolution strategy is quite similar to the previous case, when we minimized the energy consumption: if an UE would decide to offload its task ($d_{k}=1$), we need to allocate the optimal transmission rate $R_k$ for any fixed compression factor $\rho_{k}$ and device clock frequency $f_{k}^{d}$. Also in this case we can obtain the optimal rate in closed form by duality theory
\begin{equation}\label{eq:R_star_max_acc}
R_{k}^*(\rho_{k},f_{k}^{d}) =
\begin{cases}
\left[\frac{B_{k}}{\ln(2)}\ln\left(\frac{Q_{ki}^{TX}h_{k}^{2}}{W(\rho_{k})\lambda_{k}S_{k}\ln(2)N_{0}}\right)\right]^{R^{+}_{max}}_, &\text{\hspace{-.5em}\scriptsize{$Q_{ki}^{TX} > 0$}}\\
% \\ \times \mathbbm{1}(Q_{ki}^{TX}(t) > 0).  % \text{\hspace{-.5em}\scriptsize{, $Q_{ki}^{TX} > 0$}}
% \\
0 &\text{\hspace{-.5em}\scriptsize{otherwise}}
\end{cases}
\end{equation}
\normalsize
Thus, for a possible offloading decision ($d_k=1$) we compute by \eqref{eq:R_star_max_acc} the optimal data transmission rate $R_k^*$ for each $\rho_{k} \in S_{k}$ and $f_{k}^{d} \in F_{k,d}$, and we select the optimal triple $T_{k}^{*}=(R_{k}^{*},f_{k}^{d*},\rho_{k}^{*})$ that minimizes the cost function in \eqref{eq:mu_Maras_UE_cost_function}.
Conversely, in order to evaluate the minimum cost of a local learning task at the $k$-th UE ($d_{k}=0$), we just need to exhaustively search for the pair $P_{k}^{*}=(f_{k}^{d*},\rho_{k}^{*})$ that would optimize the accuracy under the prescribed constraints. Finally, depending on which one of the two optimal allocation strategies guarantees the best accuracy, we decide to offload ($d_k=1$), or not ($d_k=0$), the $k$-th user task, using the associated optimal allocation strategy $T_{k}^{*}$, or $P_{k}^{*}$, respectively.

\subsubsection*{ES sub-problem}
From the ES perspective, the optimization problem is  similar to the mu-MEDA, except for small differences in the cost function, and is expressed by
\begin{align}
\!\min_{\Phi_{s}}       &\quad -\sum_{k=1}^{K} \sum_{i=1}^{L_{k}} \tau Q_{ki}^{ES} J_{ki}^{s}f_{ki}^{s}+ \eta O\kappa \tau f_{s}^{3} \notag
\\[0.5pt]
\text{s.t.}
&\quad 0 \leq f_{ki}^{s}(t) \leq \min\left(f_{s},\frac{Q_{ki}^{ES}}{\tau J_{ki}^{s}}\right),\notag \\
%\breackequation \quad \forall  k,t \notag \\ &
&\quad \sum_{k=1}^{K}\sum_{i=1}^{L}f_{ki}^{s} \leq f_{s},\quad f_{s} \in \mathcal{F}_{s},\qquad \forall k,t
%\end{alignat}
\label{eq:server_prob2}
\end{align}
where $\Phi_{s}=[f_{ki}^{s},f_{s}]$, and can be solved likewise the mu-MEDA formulation.

\section{Simulation Results}\label{sec:sim_res}
In this section, we present the simulation results we obtained by the two optimization strategies we proposed and solved. 
% \textcolor{red}{We have already highlighted in Section \ref{sec:Accuracy_Model} that some terms of the objective functions are handled by proper LUTs, which have been experimentally filled with the values we obtained by (off-line) training and testing all the CE-CC classification networks, each one associated to a specific compression factor $\rho \in \mathcal{S}_k$ in the CE of the $k$-th UE.
% Specifically,} 
Tables \ref{tab:deep_ae_lut_param}-\ref{tab:short_ae_lut_param} report the values of the accuracy $G_k(\rho)$, the data-units $J_{k}^{d}(\rho)$ that can be compressed (and zipped by JPEG) in a clock-cycle by the $k$-th UE, when it decides to offload the classification, and the data-units $J_{k}^{L}(\rho)$ that can be compressed and classified locally in a clock-cycle by the same UE. Table \ref{tab:common_parameters} reports the data-units $J_s(\rho)$ that can be classified in a clock-cycle at the ES, as well as the image-size $M(\rho)$ and the average number of bits/pixel $N(\rho)$ that are shared by both the short- and deep-CE, when using JPEG.

\begin{table}[ht]
    \caption{LUTs parameters}
     \centering
      \caption{Deep-CE}
       \label{tab:deep_ae_lut_param}
         % leftmost table of the top level table
     \begin{tabular}{|c|c|c|c|}
    \hline
    $\rho$ &  $G(\rho) \hspace{1pt} [\%]$ %& \textbf{$\frac{1}{J_{c}}$ $[\frac{C}{DU}]$} & \textbf{$\frac{1}{J_{zip}}$ $[\frac{C}{DU}]$}
    & \textbf{$J_{k}^{d}(\rho)$ $[\frac{DU}{C}]$} & \textbf{$J_{k}^{L}(\rho)$ $[\frac{DU}{C}]$}\\ \hline
    2  & 97.3 %& \num{4.454e6} & \num{2.5e6}
    & \num{1.44e-7}  & \num{8.35e-8} \\ \hline
    4  & 96.5 %& \num{6.46e6}  & \num{1.48e6}
    & \num{1.26e-7} & \num{9.04e-8}\\ \hline
    8  & 93.4 %& \num{7.752e6} & \num{8.74e5}
    & \num{1.16e-7} & \num{8.90e-8}\\ \hline
    16 & 91.8 %& \num{8.568e6} & \num{6.84e5}
    & \num{1.07e-7} & \num{8.73e-8} \\ \hline
    32 & 83.0 %& \num{7.41e6} & N/D
    & \num{1.35e-7} & \num{1.06e-7}\\ \hline
    64 & 67.0   %& \num{7.6e6}  & N/D
    & \num{1.32e-7} & \num{1.09e-7} \\ \hline
\end{tabular} 
\end{table}
    \begin{table}
    \centering  
    \caption{Short-CE}
    \label{tab:short_ae_lut_param} 
    \begin{tabular}{|c|c|c|c|}
    \hline
    $\rho$ &  $G(\rho) \hspace{1pt} [\%]$ %& \textbf{$\frac{1}{J_{c}}$ $[\frac{C}{DU}]$} & \textbf{$\frac{1}{J_{zip}}$ $[\frac{C}{DU}]$}
    & \textbf{$J_{d}(\rho)$ $[\frac{DU}{C}]$} & \textbf{$J_{k}^{L}(\rho)$ $[\frac{DU}{C}]$}\\ \hline
    2  & 97.3 %& \num{4.454e6} & \num{1.05e7}
    & \num{1.44e-7} & \num{8.35e-8} \\ \hline
    4  & 95.8 %& \num{4.454e6}  & \num{5.17e6}
    & \num{1.68e-7}  & \num{1.10e-7}\\ \hline
    8  & 91.5 %& \num{4.454e6} & \num{8.85e6}
    & \num{1.88e-7} & \num{1.26e-7}\\ \hline
    16 & 91.3 %& \num{4.454e6} & \num{4.93e6}
    & \num{1.95e-7} & \num{1.38e-7}\\ \hline
    32 & 77 %& \num{4.454e6} & N/D
    & \num{2.25e-7} & \num{1.55e-7}\\ \hline
    64 & 50.0   %& \num{4.454e6}  & N/D
    & \num{2.25e-7}  & \num{1.65e-7} \\ \hline
\end{tabular} 
\end{table}
\begin{table}[ht]
   \caption{Common parameters}
   \label{tab:common_parameters}
    \centering
  \begin{tabular}{|c |c |c |c |c|}
  \hline
$\rho$ & $M(\rho) \hspace{3pt} [px]$ &  $N(\rho)$ $[\frac{bits}{px}]$ %& \textbf{$\frac{1}{J_{s}}$ $[\frac{C}{DU}]$}
& \textbf{$J_{s}(\rho)$ $[\frac{DU}{C}]$} \\ \hline
2  & 128x128x3 & 1.08
%& \num{1.02e8}
& \num{1.2e-7}  \\ \hline
4  & 64x64x3   & 2.27
%& \num{3.40e7}
& \num{2.17e-7} \\ \hline
8  & 32x32x3   & 4.72
%& \num{2.04e7}
& \num{2.87e-7}  \\ \hline
16 & 16x16x3   & 9.06
%& \num{1.36e7}
& \num{3.57e-7} \\ \hline
32 & 8x8x3     & 8
%& \num{1.23e7}
& \num{5e-7} \\ \hline
64 & 4x4x3     & 8
%& \num{1.19e7}
& \num{6.25e-7}
\\
\hline
\end{tabular}
  \centering
  \caption{Channel type}
   \label{tab:channel_scenarios}
  \begin{tabular}{|c |c |c |c |c|}
  \hline
Ch. Type & \textbf{$D \hspace{3pt} [m]$} & \textbf{$B \hspace{3pt} [kHz]$} & \textbf{$f_{0}\hspace{3pt} [GHz]$} & \textbf{$\sigma_{0}^2$} \\ \hline
\bf{A}& 50& 2500& 6& $\num{1.06e-10}$  \\ \hline
\bf{B}& 500& 2500& 9& $\num{2.72e-14}$ \\ \hline
\end{tabular}
\end{table}
We assumed a flat-fading channel, whose  statistical characterization is based on the \emph{Clarke's autocorrelation function} \cite{molisch2011statistical}. We considered two operating scenarios, summarized in Table \ref{tab:channel_scenarios}, and we accordingly set the time-slot duration to $\tau=50ms$, which corresponds to the the channel coherence time. The parameter $\sigma_{0}^2$ models the wireless channel power path-loss and it has been computed by considering the \emph{Alpha-Beta-Gamma} model \cite{sun2016propagation}. In a first set of simulations we considered a scenario with $K=5$ UEs connected to the network. Although this is not strictly necessary, we assumed that the devices of all the UEs share the same computation frequency set $\mathcal{F}_{d}=\{0.1,0.2,\ldots,0.9,1\}\times 1.4 \hspace{0.1 cm} GHz$, while the server computation frequency set is $\mathcal{F}_{s}=\{0.1,0.2,\ldots,0.9,1\} \times 4.5 \hspace{0.1 cm} GHz$. Finally, for simplicity, we considered an effective switched capacitance $\kappa=\num{1.097e-27}[\frac{s}{cycles}]^3$ for all the UEs and for the ES. We underline that all the simulation results have been obtained at convergence of the tested strategies \cite{neely2010stochastic}.
% \textcolor{red}{indeed, all the stochastic algorithms considered herein are characterized by a transient phase during which the imposed constraints could be violated and whose duration can be controlled acting on the step-sizes in \eqref{eq:mu_meras_virtual_queues} and  \eqref{eq:mu_maras_virtual_queues}.}

\begin{figure}[t]
\centering
    \includegraphics[width=0.90\linewidth]{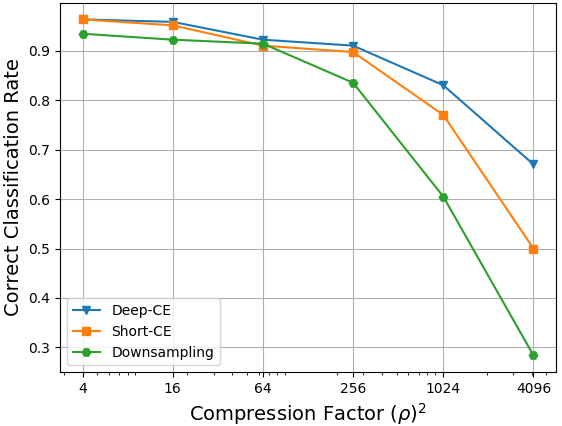}
    \caption{Classification accuracy comparison}
    \label{fig:chained_training_results}
\end{figure}

\subsection{Goal-Oriented compression results}
For simplicity, all the UEs were assigned the same image classification task, based on the German Traffic Sign Recognition Benchmarks (GTSRB) \cite{GTSRB2011} dataset. This dataset includes \text{1213} pictures of German road signals, divided in \text{43} different classes. The dataset has been split in a \text{80\%} training set, composed of \text{970} images, and \text{20\%} test set, composed of \text{243} images. During the data loading phase, all the images have been normalized to a size of \text{256x256}, and converted to a 3-channel image (one channel for each RGB color), such that the initial size of each data-unit, is \text{256x256x3}. Although this is not strictly necessary, we assumed that all the UEs share the same bank of CE-CC classification networks, e.g., the compression factors $\rho_k$ assume values on the same fixed set $\mathcal{S}=\{\text{2, 4, 8, 16, 32, 64}\}$.
In order to shade light on the performance obtained by the proposed resource managements, we find useful to show in Fig.\ref{fig:chained_training_results} the average accuracy on the test-set associated to different compressive architectures: i) Deep-CE, ii) Short-CE, iii) Down-sampling with anti-aliasing pre-filter. As expected, the accuracy $G(\rho)$ has a monotone decreasing behavior with respect to the compression factor, for all the models. The deep-CE has always the best performances even if, for lower compression factors (up to 16), the differences with the Short-CE are almost negligible. In contrast, for the highest ones (i.e., \text{32, 64}) there is a clear advantage in using the deep-CE.  For \text{compression factor} $\rho=64$ we get output tensors with a size of \text{4x4x3=48} pixels: despite (pseudo) images of this size have clearly undergone a heavy transformation, the deep-CE still allows the ES's CC to classify them with a \text{67\%} accuracy, which is still a remarkable performance for a 43-class classification task. Conversely, for this compression factor neither the down-sampling strategy nor the short-CE, allow a meaningful classification. The price to be paid for an increased accuracy of the deep-CE is the increase of the computation energy and processing delay (as summarized in Tables \ref{tab:deep_ae_lut_param}-\ref{tab:short_ae_lut_param}) that we trade by our resource management policies. 

% We remark that the decimation strategy has the worse learning performances for all the possible compression factors: however, since its very low computational cost, it makes sense to consider it in an optimal resource allocation perspective.

% \textcolor{red}{Summarising, these first set of simulations confirm, as expected and desired, that even in the absence of any communication strategy, i.e., without any ES offloading, the classification performance granted by the CEs compression scheme we considered, are always better than those of the simpler down-sampling compression, especially for the highest compression factors.}

\subsection{mu-MEDA results}
First of all, we tested the mu-MEDA strategy comparing the CE (short and deep) with the down-sampling compression strategy in channel scenario $B$, reported in Table \ref{tab:channel_scenarios}. We set the same {\it latency} constraint $D_{k}^{avg} = Q_{k}^{avg}/\overline{A}_{k} = 0.20\;s$, for all the UEs. We considered a task arrival process with $\overline{A}_{k}=2 DU/slot$, and we forced the UEs to always offload the classification task to the ES, without any opportunistic strategy (i.e., $d_{k}(t)=1, \; \forall k,t$).

\begin{figure}[t]
\begin{minipage}[b]{1.00\linewidth}
\includegraphics[width=0.90\linewidth]{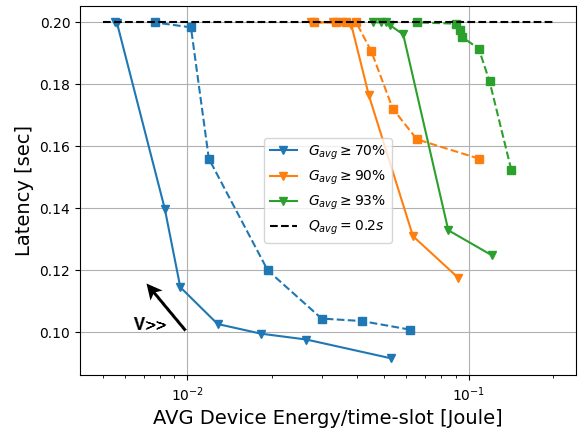}
    \centering
    \caption{UE Energy/Latency trade-off. CE (solid) vs down-sampling (dashed).}
    \label{fig:mu_meras_device_to_comparison}
\end{minipage}
\hfill
\quad
\begin{minipage}[b]{1.00\linewidth}
%\centering
    \includegraphics[width=0.90\linewidth]{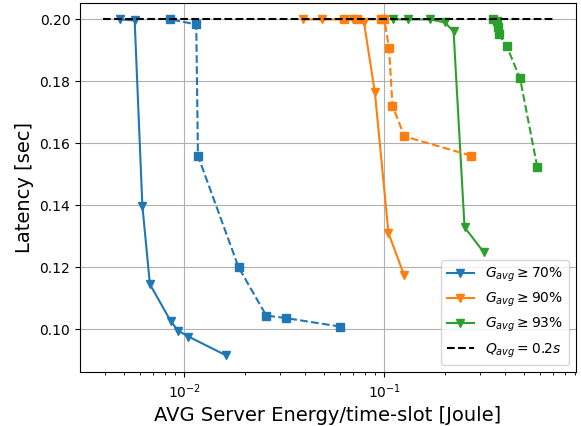}
    \centering
    \caption{ES Energy/Latency trade-off. CE (solid) vs down-sampling (dashed).}
    \label{fig:mu_meras_es_to_comparison}
\end{minipage}%
\end{figure}

Each trade-off curve in Figures \ref{fig:mu_meras_device_to_comparison} and \ref{fig:mu_meras_es_to_comparison} is associated to a different accuracy constraint, while they all respect the same latency constraint, which is highlighted by a dashed horizontal line in the plot. Each curve is obtained by evaluating the solution (at convergence) of the resource optimization problem, for several different values of the trade-off parameter $V$ in \eqref{eq:lyapunov_drift_plus_pentalty}. Specifically, by increasing $V$ we end-up to solutions characterized by a lower energy consumption and a higher latency and, as indicated by the black arrow on the figures, we move from the bottom-right to the top-left corner of the trade-off plots, which correspond to the desired optimal solutions on the borders of the feasibility regions. Figure \ref{fig:mu_meras_device_to_comparison} shows that, from the UE's perspective, there is a clear advantage on employing the CE compression strategy, since we end-up to solutions characterized by a lower (computational and transmission) energy consumption, while satisfying the same latency and accuracy constraints. This depends on the fact that {\it channel-B} is characterized by a huge attenuation: thus, since the CE compression strategy allows to satisfy the same accuracy constraint transmitting smaller DUs with respect to classical down-sampling, this allows to reduce the transmission energy expenditure considerably, without spending too much in extra computational energy for CE-based compression at the UE. Actually, the proposed dynamical, goal-oriented, compression strategy leads also to a lower ES's energy computational expenditure, as witnessed from Fig.\ref{fig:mu_meras_es_to_comparison}. Indeed, also the classification of smaller DUs is cheaper from a computational and energetic perspective.

\begin{figure}[t]
\begin{minipage}[b]{1.00\linewidth}
\centering
\includegraphics[width=0.90\linewidth]{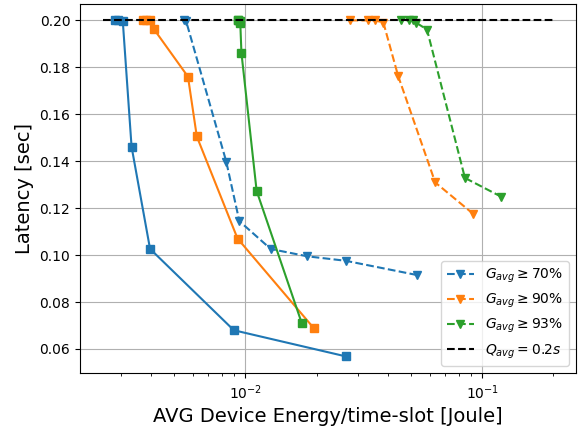}
    \caption{UE's energy/latency trade-off. Opportunistic offloading (solid) vs only offloading strategy (dashed).}
    \label{fig:opportunistic_offloading_comparison}
\end{minipage}
\hfill
\quad
\begin{minipage}[b]{1.00\linewidth}
\centering
    \includegraphics[width=0.90\linewidth]{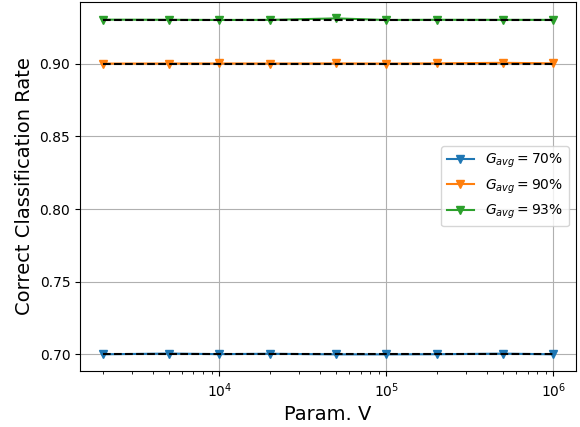}
    \caption{Average Accuracy vs $V$ with opportunistic offloading at convergence}
    \label{fig:accuracy_v}
\end{minipage}%
\end{figure}

%\centering
\begin{figure}[ht]
\centering
   \includegraphics[width=0.90\linewidth]{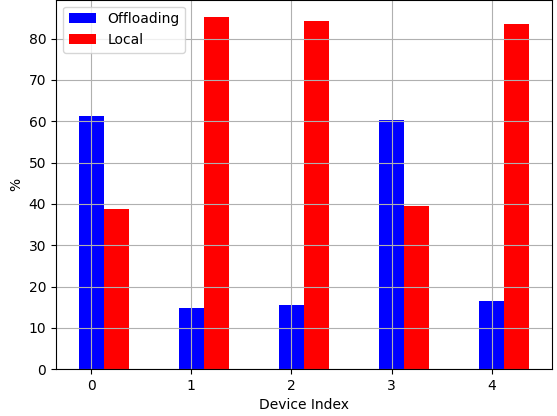}
    \captionof{figure}{\% of Offloading ($G_{k}^{avg}=70\% \hspace{1pt} \forall k$, $V=\num{1e6}$).}
    \label{fig:offloading_decision_hist}
\end{figure}
\begin{table}[ht]
    \centering
      \begin{tabular}{|c |c |c|}
      \hline
    UE & \textbf{Ch. Type} & $\kappa [\frac{s}{cycles}]^3 $  \\ \hline
    \bf{0} & \bf{A}& $10 \times \kappa_{0}$ \\ \hline
    \bf{1}  & \bf{A}&  $20 \times \kappa_{0}$ \\ \hline
    \bf{2} & \bf{B}& $30 \times \kappa_{0}$\\ \hline
    \end{tabular}
     \captionof{table}{Simulation scenarios for each UE}
    \label{tab:ue_conditions}
\end{table}

\subsection{Opportunistic Offloading}
We compared the previous scenario, where UEs always offload decision tasks to the ES, with the opportunistic offloading strategy where UEs can also decide to perform classification locally, by the same CE-CC classification architecture. Specifically, two out of five UEs are connected to the ES by the channel in {\it scenario A} of Table \ref{tab:channel_scenarios}, while the other ones by the channel in {\it scenario B}.
The opportunistic offloading strategy ends up to a dynamical resource optimization that is characterized by a significant lower UE energy expenditure with respect to the {\it always offload} strategy, still satisfying both the accuracy and latency constraints, as shown by Figs.\ref{fig:opportunistic_offloading_comparison}-\ref{fig:accuracy_v}, where clearly all the solid curves are on the left, e.g., with a lower energy expenditure, with respect to the dashed curves of the pure offloading strategy.
Figure \ref{fig:offloading_decision_hist} shows the histogram of the offloading decisions for each UE, for a (minimum) accuracy constraint $G_{avg}=70\%$ and a trade-off parameter $V=\num{1e6}$.
As expected, since the UE-$0$ and UE-$3$ experience good channel conditions, they decide to offload more frequently than the other devices, whose Channel-B requests much higher transmission power to allocate rates to the UEs and, sometimes, it may be also unfeasible to respect either the accuracy or the delay constraint, or both.

\begin{figure*}[ht]

\minipage{0.32\textwidth}
  \includegraphics[width=\linewidth]{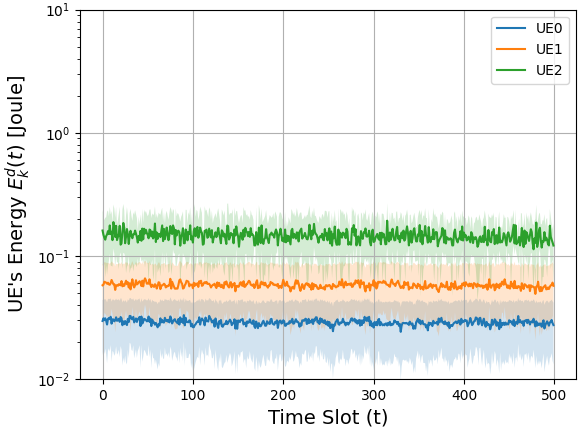}
\endminipage\hfill
\minipage{0.32\textwidth}
  \includegraphics[width=\linewidth]{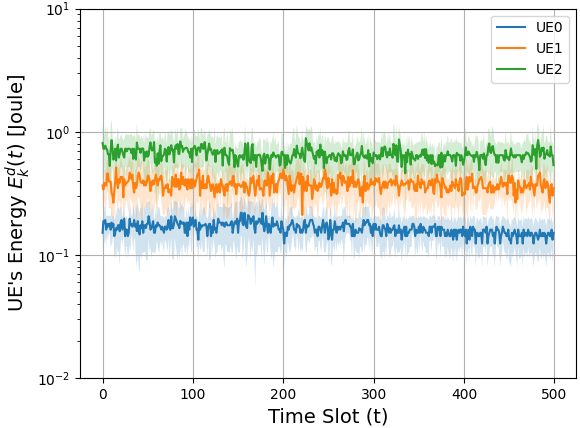}
\endminipage\hfill
\minipage{0.32\textwidth}%
  \includegraphics[width=\linewidth]{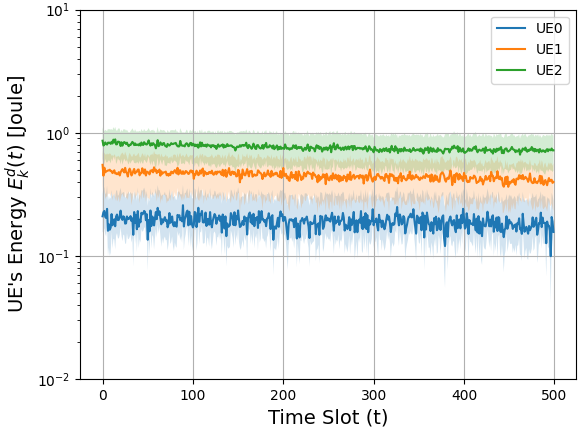}
\endminipage
\caption{Instantaneous UEs consumption for the dynamic optimization (a), fixed accuracy (b) and fixed rate (c).}
\label{fig:inst_energy_consumption}
\end{figure*}

\subsection{Comparison with static allocation strategies}
A key strength of the proposed approach is the joint \textit{dynamic} optimization of transmission\&computational resources, together with the optimal dynamic selection of the classification architecture used to perform the task. Thus, we compare the proposed multi-user optimization strategy with:
\begin{itemize}
\item A \textit{Fixed-Accuracy} optimization strategy, where we optimize both the computational and the transmission resources at the UE-side, by keeping fixed a single CE-CC classification architecture. This approach is quite similar to the one presented in \cite{merluzzi2021wireless}.
\item A \textit{Hybrid} static/dynamic optimization strategy, where, inspired by \cite{liu2022adaptable}, we fix the transmission rate $R$ on the basis of the average channel conditions, while we dynamically optimize the CE-CC architecture, as well as the computational resources at the UEs. The transmission rate $R$ is fixed as the minimum one that guarantees the stability of the UE queue. This rate can be computed through the capacity for flat-fading Rayleigh channels (eq. (9) in \cite{li2005rayleigh}), and it fixes also the transmission power.
\end{itemize}
In this case we considered a scenario with $K=3$ UEs, each one experiencing different channel conditions and computational efficiency, as summarized in Table \ref{tab:ue_conditions}. We set an arrival task with $\overline{A}=2DU/slot$, and we imposed the same accuracy and latency constraints for all the UEs to $G_{k}^{avg}=92\%$ and $D_{k}^{avg}=0.2s$, respectively. Thus, for the \textit{Fixed-Accuracy} optimization strategy, we considered the short-CE with $\rho_{k}=8$ as the unique learning model, which according to Table \ref{tab:short_ae_lut_param} is capable to grant the requested average classification performance with a fairly moderate computational energy. Figure \ref{fig:inst_energy_consumption} shows that employing a fully dynamic optimization strategy leads to solution characterized by a lower UE energy consumption. As expected UE-0 and UE-2 reach the lowest and highest energy consumption, respectively, given their computational and channel conditions summarized in Tab. \ref{tab:ue_conditions}.
It is clear that, for all the UEs, our optimization strategy allow to reach the lowest energy consumption, thus confirming the effectiveness to \textit{jointly and dynamically} optimize the transmission/computation resources as well as the learning architecture (i.e., the pair of CE-CC) to be employed, depending on the instantaneous system conditions.

\begin{figure}[ht]
\centering
\includegraphics[width=0.90\linewidth]{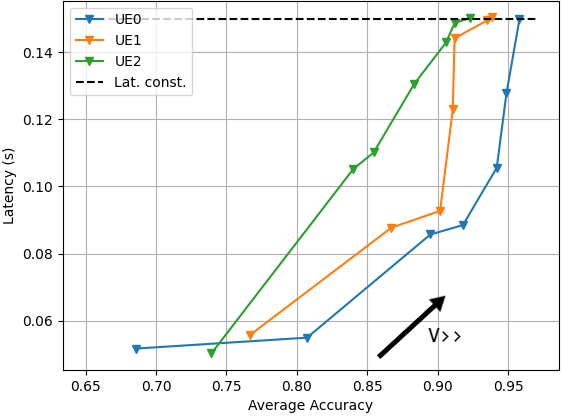}
    \caption{Accuracy vs Latency trade-off.}
    \label{fig:acc_latency_trade_off}
\end{figure}
\begin{figure}[ht]
\centering
 \includegraphics[width=0.90\linewidth]{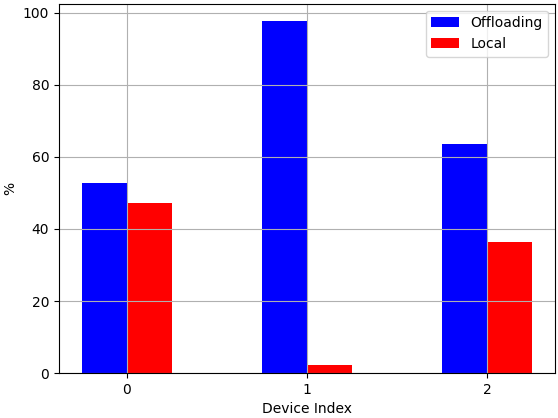}
    \caption{Offloading histograms ($V=\num{1e5}$).}
    \label{fig:offloading_decisions_hist_MA}
\end{figure}

\subsection{mu-MADE results}
We tested the mu-MADE optimization strategy considering a scenario with $K=3$ UEs, each one characterized by different channel and computational conditions. In particular, we considered an effective switched capacitance $\kappa_{0}=\num{1.097e-27}[\frac{s}{cycles}]^3$ for the ES, and higher values for the UEs, in order to simulate a lower energetic efficiency. The UE energy constraint has been set to $E_{k}^{avg}=\num{128e-3}J$. Table \ref{tab:ue_conditions} summarizes the different conditions for the devices considered in the simulation, where we employed, concurrently, both Deep- and the Short-CE. We remark that UE-0 experiences both good channel conditions and computational efficiency: this means that it has the maximum degree of flexibility on the management of the opportunistic offloading. UE-1 is characterized by the same channel conditions of UE-0, with a lower computational efficiency, while UE-2 operates with both a bad  channel and a low computational energy efficiency.

The curves shown in Fig. \ref{fig:acc_latency_trade_off} represent the accuracy-latency trade-off: by increasing the parameter $V$ of \eqref{eq:mu_meras_lyap_drift_plus_penalty}, we end up with solutions with higher accuracy and latency, moving on the curves from bottom-left to top-right corner, where we get the desired optimal solutions at the boundary of the decision region. Specifically, Fig. \ref{fig:acc_latency_trade_off} shows that UE-0 (i.e., the UE with the best computational \& channel conditions) gets the highest accuracy, while widely satisfying the latency constraint. We note a similar behaviour for UE-1 and UE-2, with a higher degree of latency for UE-2 (i.e., the device that works in the worst conditions).
Finally, we report in Fig. \ref{fig:offloading_decisions_hist_MA} the histogram of the offloading decisions for each UE. Given its favorable channel and computational energy efficiency, we have a balanced situation for UE-0, since it has the highest flexibility to choose if offloading computations, or not. On the other hand, UE-1 mostly performs offloading, since the transmission of DUs in a channel with fairly low attenuation allows to mitigate the burden due to the low computational energy efficiency. Finally UE-2, although it has a much worse channel, it offloads more DUs than UE-0 s due to its much higher computational inefficiency.

% \subsubsection*{Ensemble of compression schemes}
% In this set of simulation we test an ensemble model composed of both the down-sampling compression strategies and the Deep and Short CEs. Each UE connected to the network has the same energy and latency constraints and experiences the channel scenario B in Table \ref{tab:channel_scenarios}. In particular, we made a comparison between the Ensemble model and the decimation compression strategy. We do not consider the opportunistic offloading, so $d_{k}(t)=1, \forall k=1,t$. The results have been obtained considering a large value of the parameter $V$, i.e., $V=\num{1e7}$ which, according to \cite{neely2010stochastic}, allows to obtain a solution very close to the optimum.
% Figure \ref{fig:tradeoff_ensemble} shows the trade-off between energy consumption and classification accuracy and it witnesses how the ensemble compression strategy allows to reach the greatest accuracy levels by respecting the same energy constraint: indeed, looking at the Figure \ref{fig:chained_training_results}, it is clear that the decimation compression strategy is the worst from the learning performances perspective. However, the obtained trade-off curves show also that the decimation compression strategy allows to widely satisfy the energy constraints of the UEs: this behaviour is highlighted by Fig. \ref{fig:energy_vs_constraints}, where we show the average per-UE energy consumption with respect to the imposed constraints.

\section{Conclusions and future directions}\label{sec:conclusion_future}
In this work we implemented a goal-oriented compression architecture based on CEs, which is exploited by two distinct dynamic optimization strategies in order to either minimize the energy consumption or to maximize the learning accuracy in a multi-user scenario, where the UEs can opportunistically decide whether and when to offload the computations toward the ES.
The extensive simulation results confirmed the effectiveness and the flexibility of the proposed approaches in different scenarios.
However, we remark that the proposed goal-oriented communication architecture, and the associated resource management strategy, could exploit also classification or learning-oriented compression strategies, that may be different from the CE-based solutions presented herein.
Future research directions include the extension to multi-server scenarios, cooperative learning tasks (e.g., Federated Learning), as well as to explicitly take into account also the battery level of each UE, which may be equipped by some energy harvesting mechanism or batteries recharge plan.
%section*{Acknowledgments}
%This work was supported by MIUR under the PRIN 2017 Liquid Edge Contract.

%{\appendix[Proof of the Zonklar Equations]
%Use $\backslash${\tt{appendix}} if you have a single appendix:
%Do not use $\backslash${\tt{section}} anymore after $\backslash${\tt{appendix}}, only $\backslash${\tt{section*}}.
%If you have multiple appendixes use $\backslash${\tt{appendices}} then use $\backslash${\tt{section}} to start each appendix.
%You must declare a $\backslash${\tt{section}} before using any $\backslash${\tt{subsection}} or using $\backslash${\tt{label}} ($\backslash${\tt{appendices}} by itself
% starts a section numbered zero.)}

{\appendices
\section{Mathematical derivations for MU-mEDA}
Two Lemmas in \cite{neely2010stochastic} are useful to solve the proposed resource optimization strategies.
\begin{lemma}\label{lemma_a}
Given a queue that evolves according to $X(t+1)=max(0,X(t)+x(t+1)-\overline{x})$, by defining $\Delta_{x}=\frac{X(t+1)^2-X(t)^2}{2}$, it is always true that $\Delta_{x} \leq \frac{(x(t+1)-\overline{x})^2}{2}+X(t)x(t+1)-X(t)\overline{x}$.
\end{lemma}
\begin{lemma}\label{lemma_b} 
The following inequality holds true:
$$\quad (\max(0,Q-b)+A)^2 \leq Q^2+A^2+b^2+2Q(A-b).$$
\end{lemma}
%\section{Mathematical derivations for MU-mEDA}
Employing Lemma \ref{lemma_a}, and recalling that, given $x \in \mathbb{R}^k$, $(\sum_{k=1}^{K}x_k)^2 \leq K\sum_{k=1}^{K}x_k^2$, for the Latency Virtual Queue $Z_{k}(t)$ we have
\begin{equation}
\begin{split}
    \Delta_{z_{k}}(t) &\leq \frac{\mu_{k}^2(Q_{k}^{tot}(t+1)-Q_{k}^{avg})^{2}}{2}
    %equationbreack \\&
    \\
    &+\mu_{k}Z_{k}(t)(Q_{k}^{tot}(t+1)-Q_{k}^{avg})\\
    & \leq \frac{\mu_{k}^{2}L_{k}}{2}[Q_{k}^{UE}(t+1)^{2}+\sum_{i=1}^{L_{k}}p_{i}Q_{ki}^{ES}(t+1)^{2}]\\
    &+\mu_{k}Z_{k}(t)(Q_{k}^{tot}(t+1)-Q_{k}^{avg})+\frac{\mu_{k}^{2}L_{k}}{2}(Q_{k}^{avg})^2, \notag
\end{split}
\end{equation}
Now, recalling \eqref{eq:device_queue}, \eqref{eq:server_queue} and using Lemma \ref{lemma_b} we can derive the following inequality

\begin{equation}
\begin{split}
    \Delta_{z_{k}}(t) &\leq \frac{\mu_{k}^2L_{k}}{2}\{Q_{k}^{UE}(t)^{2}+M_{k}^{UE}+2Q_{k}^{UE}(t)(A_{k}(t)\\
    &-N_{k}^{UE}(t))+\sum_{i=1}^{L_{k}}[p_{i}Q_{ki}^{ES}(t)^2+M_{k}^{ES}\\
    &+2p_{i}Q_{ki}^{ES}(t)(\hat{A}_{ki}(t)-N_{ki}^{ES}(t))]\}\\
    %equationbreack \\&
    &+\mu_{k}Z_{k}(t)(Q_{k}^{tot}(t+1)-Q_{k}^{avg})+\frac{\mu_{k}^{2}L_{k}}{2}(Q_{k}^{avg})^2 \notag
\end{split}
\label{eq:bound_z}
\end{equation}

where $\hat{A}_{ki}(t)=\mathbbm{1}\{\rho_{k}(t)=s_{ki}\}N_{k}^{UE}(t)$, $M_{k}^{UE}=A_{k,max}^{2}+N_{kdev,max}^{2}$ and $M_{ik}^{ES} =A_{ki,max}^{2}+N_{ki,max}^{2}$ . 
The same derivations presented in \cite{merluzzi2020dynamic} can be applied to the accuracy virtual queue, thus obtaining an upper-bound for $\Delta_{y_{k}}(t)$. Putting together the derived instantaneous upper-bounds we end up to the optimization problem presented in  in Sec.\ref{subsec:mu-MEDA}.
% \small
% \begin{equation}
%     \Delta_{p}\{\Theta(t)\} \leq \mathbb{E}\left\{\sum_{k=1}^{N}\Delta_{z_{k}}(t)+\Delta_{y_{k}}(t)\right\}+V\mathbb{E}\left\{E_{tot}(t)\right\}.
% \end{equation}
% \normalsize

% \small
% \begin{equation}
% \begin{split}
%     \Delta_{y_{k}}(t) &\leq \frac{\nu_{k}^2(G_{k}^{avg}-G_{k}(t))^{2}}{2}+\nu_{k}Y_{k}(t)(G_{k}^{avg}-G_{k}(t))\leq \frac{\nu_{k}^{2}}{2}(G_{k}^{avg}-G_{k}^{*})+\nu_{k}Y_{k}(t)(G_{k}^{avg}-G_{k}(t)),
% \end{split}
% \label{eq:bound_y}
% \end{equation}
% \normalsize
% where $G^{*}$ is either the maximum or minimum accuracy $G_{max}$ and $G_{min}$, respectively, depending which one is the farthest from $G_{avg}$. 
% Since the optimization variables affect only the terms $N_{k}^{UE}(t)$, $N_{ki}^{ES}(t)$ and $G_{k}(t)$, all the terms which do not depend on them can be neglected. Using the previous inequalities we can bound the Lyapunov Drift plus penalty function
% \small
% \begin{equation}
%     \Delta_{p}\{\Theta(t)\} \leq \mathbb{E}\left\{\sum_{k=1}^{N}\Delta_{z_{k}}(t)+\Delta_{y_{k}}(t)\right\}+V\mathbb{E}\left\{E_{tot}(t)\right\}.
% \end{equation}
% \normalsize
%  Furthermore, according to {\it stochastic optimization} theory we can remove the expectations, thus obtaining the optimization problems presented in Secs.\ref{subsec:mu-MEDA}-\ref{subsec:mu-MADE}.
}

\balance

\bibliographystyle{IEEEtran}
\bibliography{EML_Multiuser}

\vfill
\end{document}